\documentclass[journal]{IEEEtran}
\usepackage{amsmath, amsfonts}
\usepackage{indentfirst}
\usepackage{comment}
\usepackage{subfigure}
\usepackage{caption}
\usepackage{graphicx}
\usepackage[table]{xcolor}
\usepackage{soul}
\usepackage{amsmath}
\usepackage{float}
\usepackage{array}
\usepackage{makecell}
\usepackage{enumitem}
\usepackage{textcase}
\usepackage{algorithm}
\usepackage{algorithmicx} 
\usepackage{algcompatible}
\usepackage[]{algpseudocode}
\usepackage{adjustbox}
\usepackage{multirow}
\usepackage{xcolor}
\def\BibTeX{{\rm B\kern-.05em{\sc i\kern-.025em b}\kern-.08em
    T\kern-.1667em\lower.7ex\hbox{E}\kern-.125emX}}
\def\baselinestretch{0.96}
\ifCLASSINFOpdf
\else
\fi
\algnewcommand{\algorithmicand}{\textbf{ and }}
\algnewcommand{\algorithmicor}{\textbf{ or }}
\algnewcommand{\OR}{\algorithmicor}
\algnewcommand{\AND}{\algorithmicand}
\algnewcommand{\var}{\texttt}
\newcommand{\mathleft}{\@fleqntrue\@mathmargin3pt}
\newcommand{\mathcenter}{\@fleqnfalse}
\begin{document}
\title{An Intelligent Resource Reservation for Crowdsourced Live Video Streaming Applications in Geo-Distributed Cloud Environment}

\author{
E. Baccour$^\ddagger$, F. Haouari$^\ast$,
A. Erbad$^\ddagger$, A. Mohamed$^\ast$, K. Bilal$^\star$, M. Guizani$^\ast$, and M. Hamdi$^\ddagger$
\thanks{$\ddagger$Division of Information and Computing Technology, College of Science and Engineering, Hamad Bin Khalifa University, Qatar Foundation.}
\thanks{$\ast$CSE department, College of Engineering, Qatar University.}
\thanks{$\star$Comsats Institute of Information Technology, Abbottabad, Pakistan.}}
\maketitle

\begin{abstract}
Crowdsourced live video streaming (livecast) services such as Facebook Live, YouNow, Douyu and Twitch are gaining more momentum recently. Allocating the limited resources in a cost-effective manner while maximizing the Quality of Service (QoS) through real-time delivery and the provision of the appropriate representations for all viewers is a challenging problem.  In our paper, we introduce a machine-learning based predictive resource allocation framework for geo-distributed cloud sites, considering the delay and quality constraints to guarantee the maximum QoS for viewers and the minimum cost for content providers. First, we present an offline optimization that decides the required transcoding resources in distributed regions near the viewers with a trade-off between the QoS and the overall cost. Second, we use machine learning to build forecasting models that proactively predict the approximate transcoding resources to be reserved at each cloud site ahead of time. Finally, we develop a Greedy Nearest and Cheapest algorithm (GNCA) to perform the resource allocation of real-time broadcasted videos on the rented resources. Extensive simulations have shown that GNCA outperforms the state-of-the art resource allocation approaches for crowdsourced live streaming by achieving more than 20\% gain in terms of system cost while serving the viewers with relatively lower latency.
\end{abstract}
\begin{IEEEkeywords}
Resource Allocation, Cloud Computing, Live Streaming, Machine Learning.
\end{IEEEkeywords}
\IEEEpeerreviewmaketitle

\section{Introduction}
\IEEEPARstart{R}{ecently} crowdsourced live streaming applications depicted massive increase in popularity. This is attributed to the technological advancement in mobile devices for live videos generation and broadcast. A prominent example is Facebook Live, which is one of the most popular social media platforms that offer live streaming services. In the first quarter of 2020, Facebook reached 2.6 billion monthly active users, which constitutes more than 30\% of internauts \cite{Facebook2}\cite{Facebook3}. Moreover, this remarkably popular platform achieved on average over 8 billion daily video views in 2019, where 20\% of  published videos are live \cite{Facebook3}.

The advancement and expansion of crowdsourced live streaming platforms encouraged the growing number of users to broadcast their own videos. However, most of the boradcasters are amateurs and their behavior is highly dynamic over time, as they can start and end their streamings randomly, which makes estimating the resource requirements hard. Furthermore, livecast platforms started to implement rich interactions between broadcasters and their viewers such as allowing live chat while watching, posting comments, adding reactions, and even donating money to support the popular content creators. Under these scenarios, live services become very sensitive to latency. Finally, the  new  generation  of  broadcasters and viewers are extremely heterogeneous in terms of device capacities and network conditions, in addition to their high  geo-distribution. This massive number of device configurations and contents deployed in the same interactive platform creates a strong need to transcode the original video version to several standard bitrates in order to serve the users with the proper qualities adequate to their capacities. For instance, Twitch.Tv is broadcasting live contents coming from more than 100 countries with more than 150 different qualities \cite{he2016coping}. As a result, satisfying viewers becomes significantly expensive.

These live streaming requirements brought a major dilemma to content providers: How to maximize the QoS while minimizing the system operational cost. More specifically, two factors contribute to increase the satisfaction of viewers. First, the perceived delay to deliver the live stream has been highlighted as the main key to increase the fidelity of users toward the crowdsourcing platform \cite{krishnan2013video}.  Experiencing stalls and startup delays or failing to interact with the broadcaster may lead the viewer to abandon the video, refrain him/her from revisiting the livecast website and force the broadcaster to publish the content in a better platform. Second, as interactive users have different capacities, the content provider should offer the maximum of bitrates to match the heterogeneous quality preferences of viewers. Both of the aforementioned criteria require massive computational
demands at the proximity of end-users, resulting in higher cost for the content provider.

Geo-distributed cloud computing has been introduced as a solution  to perform elastic and cost-effective transcoding due to its 'pay as you go' feature \cite{wei2017qos}. Particularly,  crowdsourcing live videos services can benefit from the distributed and on-demand renting of computational resources, without making any commitment or upfront payment \cite{he2014cost}. Recently, various researchers considered on-demand cloud provisioning to optimize  the transcoding and delivery latency in a geo-distributed environment. Authors in \cite{he2016coping} and  \cite {bilal2018qoe}  designed a dynamic resource scheduling and  allocation based on content popularity, aiming at enhancing the viewers satisfaction. Meanwhile, the work in \cite{dong2019joint} was devoted to accommodate live video services in geo-distributed clouds  while  considering  the  satisfaction  of both  broadcasters and viewers. 

The above-mentioned works considered only on-demand renting of cloud instances when receiving live streams. However, renting on-the-fly suffers form two major challenges: (a) live crowdsourcing applications require strict streaming and startup delays,  whereas, a cloud instance takes considerable time to activate and become functional. Authors in \cite{he2016coping} experienced Amazon EC2 cloud renting and reported that it takes up to two minutes to boot up servers and start running the tasks, (b) on demand procurement is very expensive  compared to pre-rented resources. Therefore, to minimize the initialization delay and renting cost, cloud resources should be pre-reserved for future usage with an upfront payment \cite{he2014cost}.
Many cloud service providers, including Amazon present up to 75\% cost reduction if computational resources are proactively reserved, which is not the case of the on-demand renting that is charged with instant pricing \cite{amazonDiscount}. Still, proactive renting presents some issues to estimate the required number of cloud instances, so that the resources should neither be over-provisioned nor under-provisioned. Indeed,  over-provisioning  can  lead  to  additional charges, while under-provisioning may result in lower serving efficiency and incapacity to match all viewers quality preferences. Furthermore, as broadcasters and viewers are characterized by their wide distribution, the location of reserved resources should be thoroughly selected.

To the best of our knowledge, little efforts touched upon the prediction and pre-renting of cloud instances for live videos. Specifically, most of the works opt to reserve resources with static and centralized strategies \cite{7508389}, which leads to possible under-provisioning and non-matching to the buffering delays expected by distributed users. Few works proposed to proactively forecast resources for cloud crowdsourced live video streaming, including \cite{prediction1,prediction2,prediction3,prediction4}. These efforts predict the future load based on the historical traffic or characteristics of past videos. Meaning, they considered only features of the incoming load and ignored to observe the characteristics of the viewers and their preferences (e.g., location, network quality, and device capacities.), which highly impacts the number of required transcoding resources. Furthermore, the authors did not take into consideration the requirements of the streaming application (e.g., latency threshold) neither the requirements of the crowdsourcing platform (e.g., cost). Finally, the previous works predicted only the total incoming traffic and did not study the incoming requests of viewers in different geo-distributed cloud sites, which impacts the serving latency. To summarize, predicting the live videos traffic load or videos characteristics helps to estimate the number of needed resources. Still, under-provisioning or over-provisioning scenarios can occur due to the heterogeneous capacities of devices of viewers incoming from all over the world and requiring different transcoding tasks. Also, the prediction of load of videos does not include any direct insight about the potential location of viewers, the QoS requirements of the application, or the budget of the crowdsourcing platform.

The novelty in our approach is to perform proactive resource reservation based on learning of past optimal allocations for geo-distributed cloud platform, by taking into consideration multi-factors: the load of videos, their characteristics, the locations of viewers and their preferences, and the requirements of the streaming application and the crowdsourcing platform. More specifically, our approach is based on using the metadata of past incoming videos and their distributed viewership to run an offline optimization that decides the required computation resources in different cloud regions for the historical data, while minimizing the cost of the network and respecting the QoS constraints (latency and quality). The optimal allocations given by the optimizer are used to train our machine learning models to proactively forecast future computational demands at different cloud sites. Different contributions of this paper can be summarized as follows:

\begin{itemize}
\item We formulate our geo-distributed transcoding resource allocation problem as an optimization, with an objective to minimize the overall cost of the network, while respecting the tolerated latency threshold and the preferences of viewers in terms of bitrate.

\item We create our time series datasets from the decisions of the optimizer on past incoming live videos. These  datasets contain the records of the required computational resource, in different time slots, at each cloud region.

\item We adopt machine learning techniques to train our distributed time series resource forecasting models and predict future computational demands.

\item We propose a GNCA algorithm to assign live incoming videos to the predicted transcoding resources, with a goal to maximize the QoS. 

\item We conduct extensive evaluation of our system using our real-life dataset publicly available at \cite{dataset}, and we illustrate that the proposed approach can minimize the  network cost compared  to recent allocation  systems. 
\end{itemize}

This paper is organized as follows: Section \ref{related} presents an overview of existing related works in the literature. Section \ref{systemModel} introduces our system model. Our offline transcoding resource allocation optimizer, that serves to   create the time series datasets, as well as the forecasting models are presented in section \ref{section:optimizer}. We present our GNCA algorithm in section \ref{GNCA}. In section \ref{section:performance}, we conduct extensive simulation to evaluate the performance of our system and compare it to different live videos resource allocation approaches. Finally, section \ref{section:conclusion} concludes the paper and discusses possible future extensions. 
\section{Related works}\label{related}
\subsection{crowdsourced live video streaming system}
The crowdsourcing live streaming services have unique properties compared to on-demand video services \cite{BACCOUR2020102801}. First, the behavior of crowdsourcers is highly dynamic over time, as broadcasters can start and end their streamings randomly, which makes it hard to estimate the resource requirements. Second, the most critical feature of crowdsourced streaming is that a broadcaster can interact with the viewers in live, making this service very sensitive to latency. Finally, the new generation of broadcasters are extremely heterogeneous in terms of device capacities, in addition to their global geo-distribution. Thus, receiving new videos with different qualities, from different regions online makes preparing the exact required resources for transocding not evident. 
\subsection{Cloud computing-based approaches for live streaming}
Cloud computing has been introduced as a prevailing solution  to perform elastic and cost-effective transcoding for
live applications \cite{wei2017qos}. To cope with the tremendous number of broadcasted live videos and the heterogeneous capacities of interactive viewers and broadcasters, efficient resource allocation design is crucial. The work in \cite{chen2015cloud} introduced a cost-effective geo-distributed scheduling for crowdsourced live streaming. The authors studied the problem of selecting the optimal number of cloud sites for crowdsourcers and proposed a collaborative and adaptive strategy to optimize leasing cloud instances for video streams allocation. However they did not address the QoS nor considered multiple video bitrates. The work in  \cite{he2016coping} adopted a geo-distributed cloud platform for live streaming with an objective to maximize the QoS and minimize the system cost. Indeed, the authors presented a metric that quantifies the viewers' satisfaction based on the capacity of the system to offer the maximum number of qualities, yet they did not consider the perceived latency as a metric that impacts the QoS. Finally, authors in \cite{bilal2018qoe} extended the latter work and added the playback delay to quantify the viewers' satisfaction. They proposed a greedy algorithm, namely GMC, for selecting the cheapest cloud sites to allocate transcoding tasks, while being constrained by the latency and quality required scores. However, the wide distribution of viewers' locations was not considered in the above work. The aforementioned related works adopted on-demand and online renting of computational instances to allocate only the required  resources. Such strategy, that consumes a large start up delays, is not  adequate for live streaming because of the property of instant content generation and the requirement for timely actions. Moreover, higher costs are incurred for on-the-fly resource allocation. The novelty of this paper is the design of a proactive resource forecasting on a geo-distributed cloud framework that predicts the number of required computational instances in different cloud regions based on optimal allocations performed on historical data. Then, at real-time, the content provider system uses the pre-paid instances to host incoming videos, while focusing on maximizing the QoS.
\subsection{Machine learning based approaches for live streaming}
Many research works used machine learning to enhance the QoS. Particularly, most of the efforts focused on buffering and bitrate selection \cite{8727887}, while others predicted the best adaptive bitrate parameters to adapt the video quality to the viewers' preferences \cite{le2018improving}. A recent work \cite{jeon2019hybrid} proposed a popularity predictive model that adopts XGboosting and deep neural networks to forecast the number of potential views of live videos, yet authors did not estimate the distribution of views in different regions, as well as the required resources to handle the transcoding of contents. Geo-distributed popularity prediction is studied in \cite{8761591,BACCOUR2020982}, where authors instantly manage to rent on-demand servers to host the content for the potential joining viewers in different cloud sites. However, on-the-fly renting costs the system the delay of servers reboot. Few works proposed to proactively reserve resources for cloud crowdsourced live video streaming. The authors in \cite{prediction1} predicted the traffic load, in order to forecast the needed transcoding resources. Then, when live videos are received, an online algorithm is used to adjust the predicted resources based on the required delay and budget. Predicting the transcoding load without taking into consideration the preference of viewers, and the delay and cost requirements of the system may lead to over or under-provisioning, resulting in additional costs to adjust the resources. The work in \cite{prediction2} predicted the future transcoding load based on the jitter deviation estimation, whereas the work in \cite{prediction4} predicted the transcoding speed and CPU consumption in transcoding servers. However, these works focused on the preference of viewers in terms of bitrate and ignored the constraints of the application in terms of latency and the requirements of the platform in terms of cost. Moreover, all the previous works do not consider the location of viewers and the allocation of resources in a distributed architecture, which highly impacts the serving latency and the infrastructure cost. To the best of our knowledge, we are the first to forecast the required resources ahead of time in a geo-distributed cloud platform based on a real-world dataset of historical optimal allocations, while proactively respecting multi-factors including the preference of viewers (e.g., bitrate, locations), the requirements of the platform (e.g., cost), and the constraints of the streaming application (e.g., latency threshold). Then, as resources are already paid, we design a heuristic that focuses on assigning the incoming streams as close as possible to end users.

In summary, the novelties in this paper are:  (1) The formulation of a strategy that minimizes the operational cost of the network and maximizes the QoS, while taking into consideration geo-distributed viewers and their bitrate requests; (2) The forecast of computational resources proactively at each geo-distributed cloud region to minimize the on-demand system cost and latency without under or over provisioning of resources; and (3) The design of online heuristic, GNCA, to assign the reserved resources to incoming videos in real-time.
\begin{figure*}[h]
\centering
  \includegraphics[scale=0.27]{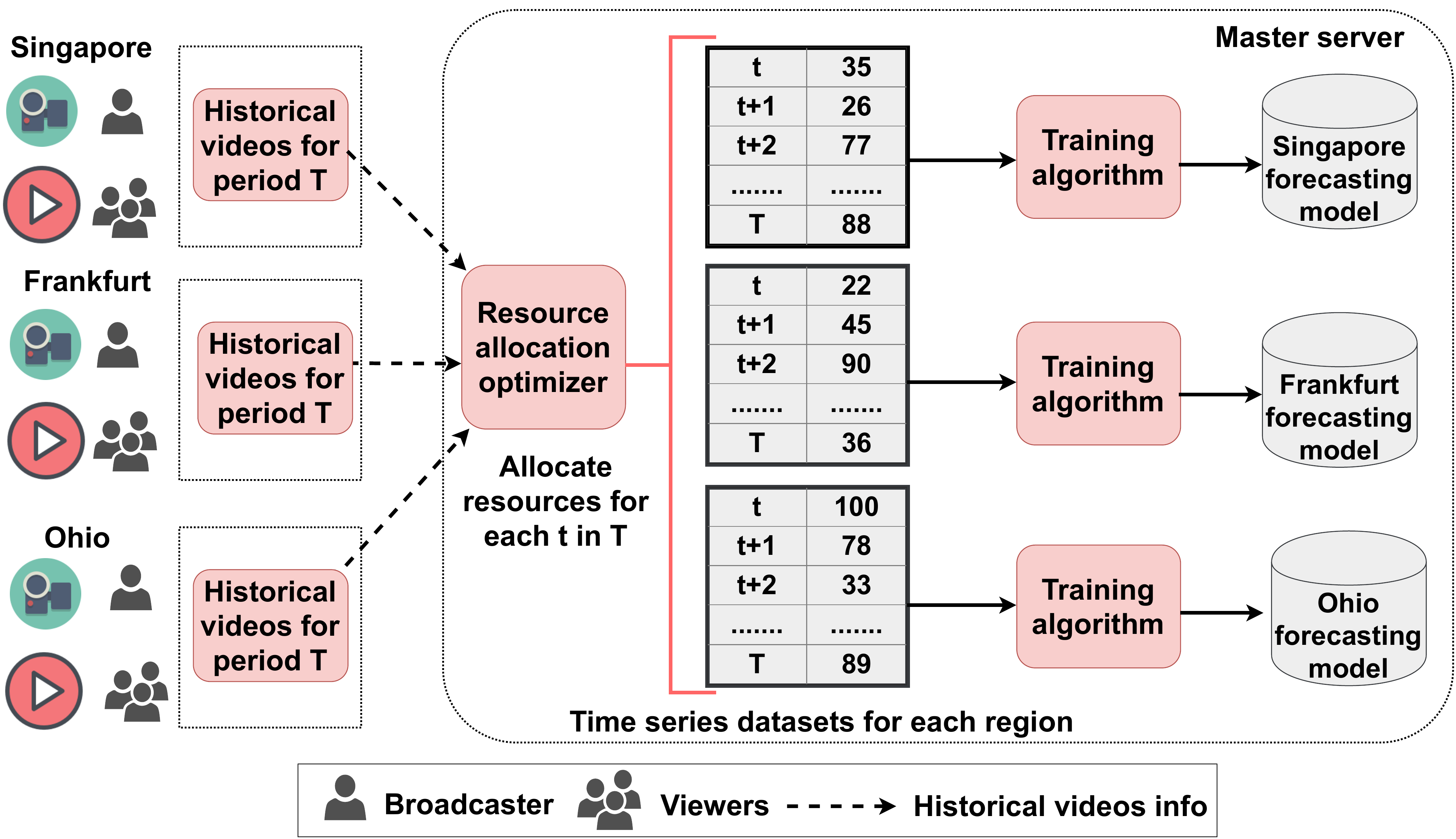}
  \caption{\small System model-offline/ Phase 1: (1) Collecting Live videos metadata, (2) Feeding the collected data to the optimizer to create datasets of required resources in each region, (3) Training the datasets to forecast future resource demand.}
  \label{fig:system model_phase1}
\end{figure*}
\section{System model}\label{systemModel}
\begin{figure*}[h]
\centering
\includegraphics[scale=0.25]{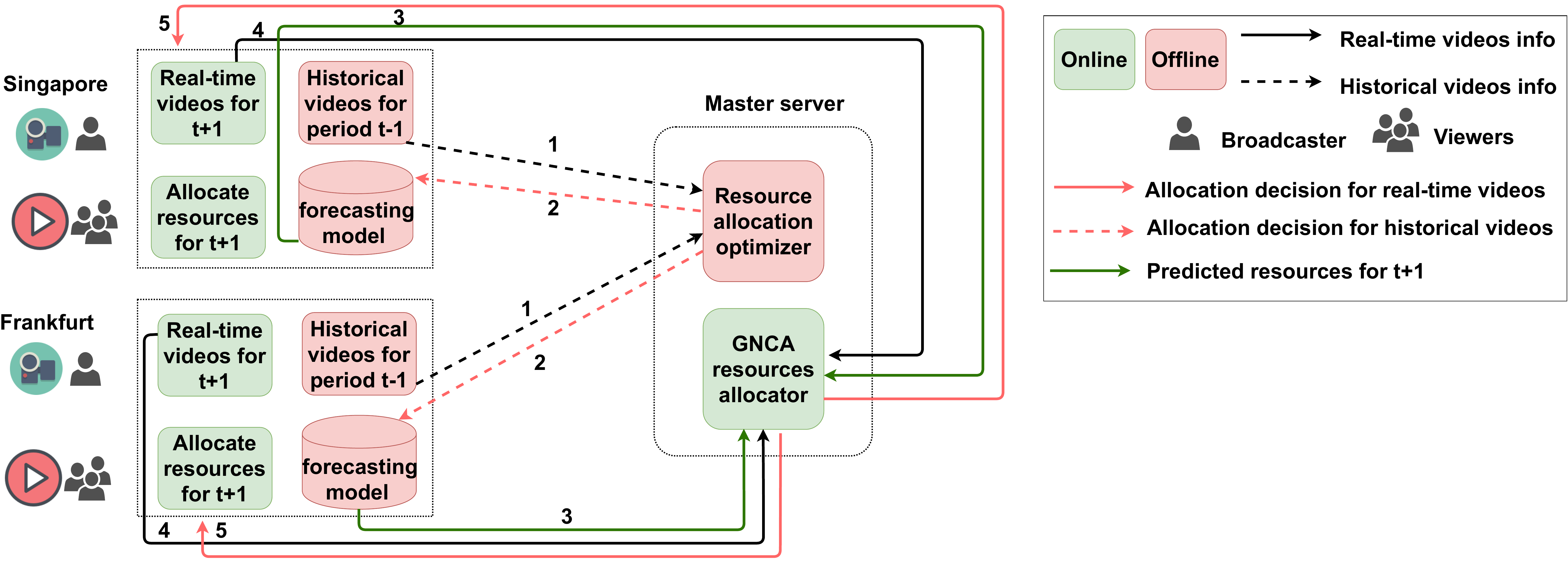}
\caption{ \small System model at real-time/Phase2:  (1) At period t, sending the collected data at (t-1) to the optimizer, (2) Feeding the allocation decisions related to each cloud site to the corresponding predictive model, (3) Notifying the GNCA resource allocator  about the forecasted resources, (4) At period (t+1), receiving real-time incoming videos, (5) Allocating the videos on the reserved resources.}
  \label{fig:system model_phase2}
\end{figure*}

In our work, a geo-distributed cloud infrastructure that comprises multiple geographically distributed data centers is adopted.  Furthermore, a centralized master server is deployed to orchestrate the system phases. More specifically, two phases are composing our framework, which are an offline phase 1 (see Fig. \ref{fig:system model_phase1}) and a real-time phase 2 (see Fig. \ref{fig:system model_phase2}).

In the offline phase 1, metadata of live videos are collected for a period T. T=\{$t_1$, $t_2$,...,$t_T$\} denotes a set of equal and consecutive time slots (e.g., hours, days, weeks, months). During these collection time periods, when geo-distributed broadcasters stream new videos, the contents along with their original bitrates, will be allocated by default in the nearest cloud region. Then, the metadata of each video, including the original quality and the viewers' locations are gathered and sent to the master server. In the master server, an optimizer is deployed. This centralized optimizer receives as an input the collected historical videos metadata and decides the optimal number of computational cloud instances required for transcoding. Moreover, the locations of rented servers across the geo-distributed data centers and the regions from where viewers should be served are also determined by the optimizer. The decisions of the optimizer serve to create, later, our  time series datasets. Each dataset is related to one cloud site and contains records of the optimal number of cloud instances that should be rented at each past slot t during T, in this site. Consequently, each cloud region gets its own dataset and predictive model that will be trained to forecast the number of cloud instances to be reserved for the upcoming time interval.

The phase 1, which is executed offline, presents an initiation to launch the online system, namely phase 2. In fact, at real-time, the trained time series models decide the required resources for future demands based on previous load of incoming videos and requests. Particularly, in the beginning of a time period t, the metadata of the incoming videos in the period (t-1), are first sent to the optimizer in the master server to get the optimal computational resources at each cloud site for the same time instance. Second, the optimizer decisions corresponding to each cloud site are fed to the related predictive model to predict the required cloud instances for the future period (t+1) and reserve them proactively. Third, the number of reserved cloud instances are loaded with our centralized GNCA algorithm to be considered when receiving the incoming videos in the fourth step. Fifth, at the period (t+1), GNCA performs the online video allocation, while taking into consideration the viewers' locations and qualities. The aforementioned steps performed in the real-time phase are presented in Fig. \ref{fig:system model_phase2}, guided by the ordered arrows.

To summarize, the first phase is executed offline to train the models that will be responsible to predict the required resources in the second online phase. When servers are  rented, the low complexity heuristic, namely GNCA, will wisely assign these resources to the live videos in real-time.
\section{Offline resource allocation optimizer and forecasting models: Phase 1}\label{section:optimizer}
\subsection{Offline optimizer}
In this section, we present the dataset that will be used in our work and different steps needed to  preprocess it. Next, we formulate the offline resource allocation as an optimization problem, that receives the processed data and decides the number and the locations of cloud servers required to transcode different live videos and serve viewers. The objective of the optimizer is to minimize the system cost, while respecting the delay and quality thresholds.
\subsubsection{Dataset}\label{section:Dataset}
In this paper, we will use the Facebook live videos dataset \cite{dataset}, collected by our team on January, February, March, May, June, and July 2018. This dataset contains more than two million video streams fetched every 3 minutes during the aforementioned collection period. In this way, each video, along with the number of viewers at the recording time, can occur in different consecutive fetches, depending on its length. In our work, the metadata of each video (e.g., number and locations of viewers) are selected from the fetch where the content reaches the maximum of its views. Other features are also collected with each stream, however,  only the broadcaster location,  the creation time,  the width and height serving to get the original bitrate and the viewers' locations are  required for the offline optimizer.
\subsubsection{Preprocessing}
The first phase of our framework presents an offline resource forecasting in the cloud, based on the demand load in each data center. Hence, as the dataset contains only raw coordinates of viewers, our data needs to be preprocessed. More specifically, we mapped the viewers to 10 different Amazon Web Services (AWS) cloud locations \cite{amazon},  which are US West-California, US East-Virginia,  US East-Ohio,  South America-Sao paulo, Europe-Paris, Europe-Frankfurt, China-Ninxgia, Asia-Singapore, Asia-Seoul, and Asia-Mumbai. In this way, each viewer will be labeled with its closer data center, until finding the number of requests incoming from different regions. Similarly, we mapped the broadcasters to their nearest cloud site. Moreover, using the height and width of videos and assuming that the amount of the motion in the image is medium and the frame rate is equal to 30 fps, we deduced the bitrate representation of all streams \cite{rhiannon_2018}. Finally, as our dataset lacks the information about the requested qualities, we used the distribution of  bandwidth capacity and screen resolutions in each continent \cite{statcounter} to classify the incoming demands for each stream into four bitrate classes: 720p, 480p, 360p, and 240p. Meaning, for each live video, we grouped the viewers into different representations, while taking into consideration that the requested quality should be lower than the original bitrate.
\subsubsection{Problem formulation}\label{section:ProblemFormulation}
We, first, define $t$ as a period of time where videos are broadcasted and metadata can be collected. Let $V(t)$=\{$v_{1}$, $v_{2}$,....$v_{m}$\} denote the set of incoming videos during the time period $t$ and let $Q$=\{$q_{1}$, $q_{2}$, $q_{3}$, $q_{4}$\} present the set of video bitrates, which is equal in our case to $Q$=\{240, 360, 480, 720\}. Additionally, we define $q^b$ as the original broadcasted bitrate, $q^{re}$ as the requested quality and $q^{tr}$ as the transcoding quality. We, also, define a binary variable $QT$ equal to 1, if $q^{re} \leq q^b$ and 0 otherwise. The set $QB(t)$=\{$q^b_1$, $q^b_2$, $q^b_3$,...,$q^b_m$\} presents the original bitrates of broadcasted videos in the time slot $t$ and the set $R$=\{$r_{1}$, $r_{2}$,....$r_{n}$\} defines the AWS regions, where $n$ is equal to 10 in our work. Let $r^b$ denote the region from where the stream is broadcasted, $r^w$ the site from where the video is served, and $r^{tr}$ the region where the video is transcoded. The round trip delay between $r^{tr}$ and $r^w$ is presented by the variable $d_{r^{tr}r^{w}}$. Furthermore, $B(t)=\{r^{b}_1,r^{b}_2,...r^{b}_m\}$ denotes the set of broadcasting regions of each incoming video during the time slot $t$. We define $P=\{P_{v_{1}},P_{v_{2}},...P_{v_{m}}\}$ as the set of viewers' distribution for each incoming video in $t$. It means, each $P_{v_{i}}$=\{$p_1$, $p_2$,....$p_n$\} denotes the set of number of viewers of the stream $v_{i}$ related to different regions. As previously described, viewers have different preferences in terms of bitrates. Hence, we define $p_r(Q)$=\{$p_r(q_1)$, $p_r(q_2)$, $p_r(q_3)$, $p_r(q_4)$\} as the distribution of viewers requesting different qualities. Finally, knowing that some videos do not have any viewer for a specific bitrate representation near some cloud sites, we introduce a binary variable $E(v,q, r^{w})$  that is equal to 1, if the bitrate $q$ of the video $v_i$ is requested from the cloud site $r^{w}$, and 0 otherwise. The video transcoding is defined as converting a higher stream quality to a lower one. In this paper, we assume that each transcoding task requires a single elastic cloud instance to be processed. For that, we will consider renting Amazon EC2 c5.large computing instances \cite{amazonDiscount}.

Next, we will introduce the optimal placement of different live videos, while taking into consideration the latency requirements of the system. The optimization is executed periodically, every $t$ frame and it relies on two decision variables. The first one is $I(v,q,r)$, which is equal to 1, if a live video $v \in V(t)$ is transcoded to the bitrate representation $q$ in the cloud region $r$, 0 otherwise. The second decision variable is $W(v,q^{re}, r^{tr}, r^{w})$, which is equal to 1, if the viewers $p_{r^w}$ at the cloud site $r^w$ are served with their requested quality $q^{re}$ from the transcoding instance at $r^{tr}$, 0 otherwise. For convenience, Table \ref{table:notations} presents some key notations used in this paper. As our objective is to minimize the expenses of the network, three operational costs should be considered: (1) rental cost $\mathbb{T}$ of transcoding cloud instances at each region; (2) migration cost $\mathbb{M}$ of the original video copy from the broadcaster site to the transcoding/streaming site; (3) serving cost $\mathbb{R}$ to transfer the data from the cloud to the viewers. The total operational cost $\mathbb{C}$ is presented as follows:
\begin{equation}\label{eq:Cost}
 \footnotesize
\begin{aligned}
\mathbb{C}=\mathbb{T}+ \mathbb{M} + \mathbb{R}.
\end{aligned}
\end{equation}
The rental cost $\mathbb{T}$ for transcoding is presented as:

\begin{equation}\label{eq:S}
\footnotesize
\begin{aligned}
\mathbb{T}=\sum_{v\in V(t)}\sum_{q^{tr}\in Q}\sum_{r^{tr}\in R}\zeta _{r^{tr}}\times I(v,q^{tr},r^{tr}).
\end{aligned}
\end{equation}
We define $\zeta _{r^{tr}}$ as the cost of renting a cloud instance per hour at the region ${r^{tr}}$. This price varies depending on the location. For example, Amazon charges 0.085\$ for renting a c5.large computing instance per hour in Ohio region, while it charges 0.131\$ for the same resource at Sao Paulo region \cite{amazonDiscount}.
\newline
The migration cost $\mathbb{M}$ is expressed as follows:

\begin{equation}\label{eq:M}
 \footnotesize
    \begin{aligned}
    \mathbb{M}=\sum_{v\in V(t)}\sum_{q^{tr}\in Q}\sum_{r^{tr}\in R}\eta _{r^{b}}\times \kappa(q^{b}) \times I(v,q^{tr},r^{tr}),
    \end{aligned}
\end{equation}
where $\eta_{r^{b}}$ is the data transfer cost per GB to migrate a copy of the stream from the broadcasting site ${r^{b}}$ to the trascoding site ${r^{tr}}$. $\kappa(q^{tr})$ denotes the size of the video with a bitrate $q^{tr}$. 
\newline
Finally, the total serving cost is presented as follows:

\begin{equation}\label{eq:R}
 \footnotesize
    \begin{aligned}
    \mathbb{R}=\sum_{v\in V(t)}\sum_{q^{re}\in Q}\sum_{r^{tr}\in R}\sum_{r^{w}\in R}\omega_{r^{tr}}\times
    \kappa(q^{re})\times p_{r^{w}}(q^{re})\times\\W(v,q^{re},r^{tr},r^{w}),
    \end{aligned}
\end{equation}
where $\omega_{r^{tr}}$ is the data transfer cost from cloud region ${r^{tr}}$ to the viewers per GB. This cost varies depending on the region and the amount of data to be transferred. As an example, in Ohio cloud data center, Amazon EC2 charges 0.09\$ if the transferred data is below 9.99TB; while it charges 0.085\$, when the data usage reaches 10 TB, and  0.05\$ when exceeding 150TB \cite{amazon}. In our paper, we assume that different charges stay constant during each studied time interval $t$.
\begin{subequations}\label{first:main}
\newline
Ultimately, our crowdsourced  live  streaming system deployed on geo-distributed cloud platform during the time interval $t$ can be formulated as follows: 
\begin{equation}
 \footnotesize
\min_{I(v,q,r) \ W(v, q^{re},r^{tr},r^{w})}  \mathbb{C}
\tag{\ref{first:main}}
\end{equation}

Subject to the following constraints:\\
\begin{enumerate}
\item [a)] Each incoming stream is hosted with its original quality by default in the adjacent data center to the broadcaster.
\begin{equation}
\footnotesize
I(v,q^{b}, r^{b})=1,  \hspace{2em}    \forall v \in V(t), \forall q^{b} \in QB(t), \forall r^{b} \in B(t)\label{first:a}
\end{equation}

\item [b)]  Serving a video $v$ with a quality $q^{re}$ from a cloud site $r^{tr}$ to viewers in the $r^{w}$ site, is not possible only if the requested version is transcoded and allocated in $r^{tr}$.
\begin{equation}
 \footnotesize
\begin{aligned}
W(v,q^{re}, r^{tr}, r^{w})\leq I(v,q^{re}, r^{tr})\\ \forall v \in V(t), \forall q^{re} \in Q, \forall r^{tr} \in R,\forall r^{w} \in R \label{first:b}
\end{aligned}
\end{equation}
\item [c)] A video $v$ is served to region $r^{w}$ with a bitrate quality $q^{re}$, only if it is requested in this region.
\begin{equation}
 \footnotesize
\begin{aligned}
W(v,q^{re}, r^{tr}, r^{w})\leq E(v,q^{re}, r^{w}) \\ \forall v \in V(t), \forall  q^{re} \in Q, \forall  r^{tr} \in R, \forall  r^{w} \in R\label{first:c}
\end{aligned}
\end{equation}

\item [d)] Viewers’ requests, if they exist, can only be handled by one data center.
\begin{equation}
 \footnotesize
\begin{aligned}
\sum_{r^{tr}\in R}W(v, q^{re},r^{tr}, r^{w})=E(v,q^{re},r^{w}) \hspace{1em}  \\
\forall v \in V(t), \forall q^{re} \in Q,\forall r^{w} \in R \end{aligned}\label{first:d}
\end{equation}

\item [e)] A video $v$ can be transcoded only to qualities lower than the original one, broadcasted by the content owner.
\begin{equation}
 \footnotesize
\begin{aligned}
\centering
I(v,q^{tr},r^{tr})\leq QT \hspace{1em} \forall v \in V(t), \forall r^{tr} \in R,\forall q^{tr} \in Q.
\end{aligned}\label{first:f}
\end{equation}

\item [f)]  The average latency to stream a video  $ v$ should respect the delay threshold $\mathbb{D}$ fixed by the crowdsourcing streaming platform. We note that the average delay to stream a live video is defined as the sum of round trip delays to serve all viewers divided by the total number of viewers.

\begin{equation}
 \footnotesize
\begin{aligned}
\frac{\sum\limits_{q^{re}\in Q}\sum\limits_{r^{tr}\in R}\sum\limits_{r^{w}\in R} p_{r^{w}}(q^{re})\times d_{r^{tr}r^{w}}\times W(v,q^{re},r^{tr},r^{w})}
{\sum\limits_{r^{w}\in R}p_{r^{w}}} \leq \mathbb{D}. 
\end{aligned}\label{first:g}
\end{equation}
\vspace{-0.5cm}
\item [g)] Decision variables are binary.
\begin{equation}
\footnotesize
\begin{aligned}
\centering
I(v,q^{tr}, r^{tr}), W(v,q^{re}, r^{tr}, r^{w}) \in \{0,1\}\label{first:i}
\end{aligned}
\end{equation}
\end{enumerate}
\end{subequations}
\begin {table}[h]

\captionsetup{font=footnotesize}
\caption {List of notations.}
\label{table:notations}
\begin {tabular}{|p{2.1cm}|p{5.9cm}|}

\hline
 \textbf{Notation} &  \textbf{Description}   \\ 
 \hline
$t$ & Period of time (e.g., hour, day, and month.)  \\
 $V(t)$ & Set of live videos received at period $t$  \\ 
 $R$ & Set of cloud sites \\ 
 $B(t)$ & Set of broadcasting regions of videos at $t$  \\ 
 $QB(t)$ & Set of original bitrates of videos at $t$  \\ 
 $r^{tr}$, $r^{w}$, $r^{b}$ & Region of transcoding, serving and broadcasting  \\ 
$P(t)$ &Set of viewers of live videos at period $t$\\
$P_v$ & Set of viewers at different $R$ for video $v$  \\
$p_r(Q)$ & Set of viewers for different bitrates of $v$ at region $r$  \\
$W(v, q^{re},r^{tr}, r^{w})$ & Binary decision variable that indicates the serving site and quality\\
$I(v, q^{tr},r^{tr})$ & Binary decision variable that indicates the allocation, transcoding site and quality\\
$E(v,q^{re},r^{w})$ & Binary variable that indicates viewers existence for a video quality\\
 $d_{r^{tr}r^{w}}$ & Round trip delay between ${r^{tr}}$ and ${r^{w}}$ \\
 $RTT$& Matrix of round trip delay between the different $R$\\ 
  $\mathbb{D}$ & Delay threshold\\
  $\kappa(q)$ & Size of a video with quality $q$\\
 $\zeta_{r^{tr}}$ & On demand cloud instance cost at region ${r^{tr}}$ \\
 $\mu_{r^{tr}}$ & Reserved cloud instance cost at region ${r^{tr}}$ \\
  $\eta_{r^{b}}$& Migration cost per GB from broadcaster region ${r^{b}}$ \\
 $\omega_{r^{tr}}$ & Serving request cost per GB from ${r^{tr}}$ \\
 $\mathbb{T}, \mathbb{M}, \mathbb{R}$  &  Transcoding, Migration, and Serving costs\\
  $\mathbb{C}$ & Overall system cost\\
  $\varepsilon$ & Forecasting models window size\\
  $DI(t)$& Set of on Demand instances at each region, at t\\
 $RI(t)$& Set of reserved instances at each region, at t\\
 CVN &Current number of viewers \\  
 DVN &Dissatisfied number of viewers \\ 
 \hline
\end {tabular}
\vspace{-0.4cm}
\end {table}
\subsection{Historical rented resources datasets}\label{timeaseriesDatasets}
\begin{figure}[h]
\centering
 \includegraphics[scale=0.5]{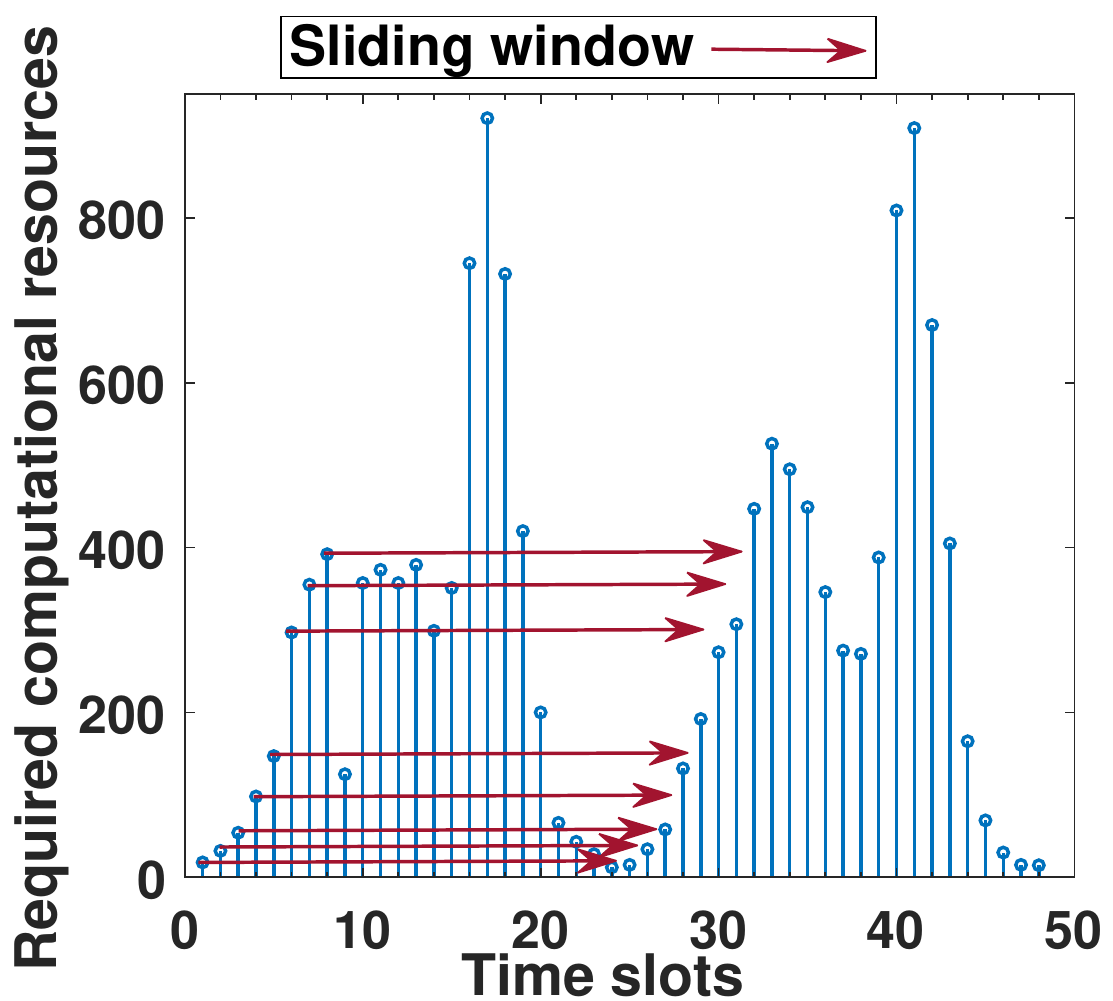}
  \caption{\small Sliding window technique applied on a sample of historical data from Singapore cloud site.} 
  \label{fig:sliding}
  \vspace{-0.5cm}
\end{figure}

In the previous section, we formulated our resource allocation problem as an optimization computed offline to decide the optimal number of computational cloud instances that should be rented at each time slot $t$, using a past collected data. The output of the optimizer is divided into $n$ datasets; where each dataset contains the records of required resources on different time slots for one of the cloud sites. We remind that $n$ presents the number of cloud regions. Finally, this time series data is restructured into  a supervised learning using the sliding window technique. More precisely, the sequence of historical $\varepsilon $ time slots related to each dataset is used as an input to the model to predict the resources to rent in the next time frame. Fig. \ref{fig:sliding} illustrates the sliding window technique, where a sequence of previous $\varepsilon $ frames (equal to 24 in the figure) is used to predict the slot indicated by the arrowhead.
\subsection{Time series forecasting models}\label{ForecastingModels}
The data in each cloud site is trained using five state-of-the art machine learning algorithms. In this work, we tried to select the techniques that proved their high performance, from different families. Therefore, we chose Long Short Term Memory (LSTM) and Gated Recurrent Unit (GRU) that belong to the Recurrent Neural Networks (RNN) widely known to be more adequate for sequential data as they are efficient in manipulating the memory state. We also selected XGboost, which is an ensemble learning algorithm known by  its  performance  compared  to  traditional  machine  learning techniques  owing  to  its  design. Finally, we tested two of the most known networks, namely MultiLayer Perceptron (MLP) which is a reference Deep Neural Network (DNN) and  Convolutional Neural Network (CNN) which belongs to Convolutional family. The mean absolute error (MAE) is used as a loss function for training. Since it is not possible to predetermine the combination of hyper parameters that gives the best results, several models are tested. Indeed, we varied the number neurons and hidden layers for GRU, LSTM, MLP and CNN and we tested different number of estimators for XGboost models. Then, based on the determination coefficient ($R^2$), we select the model giving the best accuracy. $R^2$ serves to determine the goodness of fit of different regression models. We note that a model gives accurate predictions, when $R^2$ depicted in equation \ref{eq:2} approaches 1.
\begin{equation}\label{eq:2}
 \footnotesize
R^{2}=1-\frac{\sum_{t=1}^{T}\left( A_{t}-P_{t}\right)^{2}}{\sum_{t=1}^{T}\left( A_{t}-\bar{A}\right )^{2}},\end{equation}

where T is the studied period, $A_{t}$ is the real number of resources predicted by the optimizer at the slot $t$, $P_{t}$ is the predicted number of instances for the same time slot, and $\bar{A}$ is the mean of all actually rented resources in all slots $t$.
\begin{algorithm}[!h]
\caption { Proactive resource reservation}
\label{Proactive}
\footnotesize
\begin{algorithmic}[1]
\State \footnotesize{\textbf{Input: $R$, \{$\zeta_1,...,\zeta_n$\}, 
\{$\eta_1,...,\eta_n$\}, \{$\omega_1,...,\omega_n$\}, $RTT$, $\kappa(q_1)$, $\kappa(q_2)$, $\kappa(q_3)$, $\kappa(q_4)$
}}

\For {$t \in{\{1,..,T\}}$}
\State \footnotesize{- Send metadata of videos $V(t-1)$ to the optimize as well as the}
\State \footnotesize{set of bitrates $QB(t-1)$ and the set of broadcasters $B(t-1)$.}

\State \footnotesize{- Solve the problem (\ref{first:main}) : $\min_{I(v,q^{tr},r^{tr})W(v,q^{tr},r^{tr},r^w)}$ $\mathbb{C}$}
\For {$region$ $r^{tr} \in R$}
\State \footnotesize{- Record required resources to allocate the videos $V(t-1)$.}
\State \footnotesize{- Send the recorded resources in t-1 to the forecasting model.}
\State \footnotesize{- Predict resources for period t+1 using  the forecasting model.}
\State \footnotesize{- Reserve the predicted computational instances for period t+1.}
\EndFor
\EndFor
\end{algorithmic}
\end{algorithm}
\section{Proactive Greedy Nearest Cheapest (GNCA) Resource Allocation Algorithm: Phase 2}\label{GNCA}
In this section, we present the phase 2 of our system, which is deployed in real-time.
\subsection{Proactive resource reservation}
After training different models, our system becomes ready to be implemented online. In fact, at the beginning of each time slot $t$, when the system starts to receive online incoming videos, the set of metadata of the broadcasted videos at the previous time period $t-1$ is sent to the offline optimizer in order to decide the optimal number of cloud instances required to serve the viewers with minimum delays and cost. The decisions of the optimizer for each geo-distributed cloud site are sent to the corresponding forecasting models to predict and reserve the required number of computation resources in each region for the incoming time period $t+1$. It is worth mentioning that the optimization is solved using CVX solver\footnote{http://cvxr.com/cvx/doc/solver.html} and Gurobi\footnote{https://www.gurobi.com/resource/mip-basics/} library on Matlab. We note that CVX is a strong modeling system for convex optimizations that supports multiple solvers, such as Gurobi, which is designed for Mixed Integer Programming (MIP). This justifies opting for these tools to solve our convex and integer-based optimization.

\begin{figure}[h]
\centering
 \includegraphics[scale=0.35]{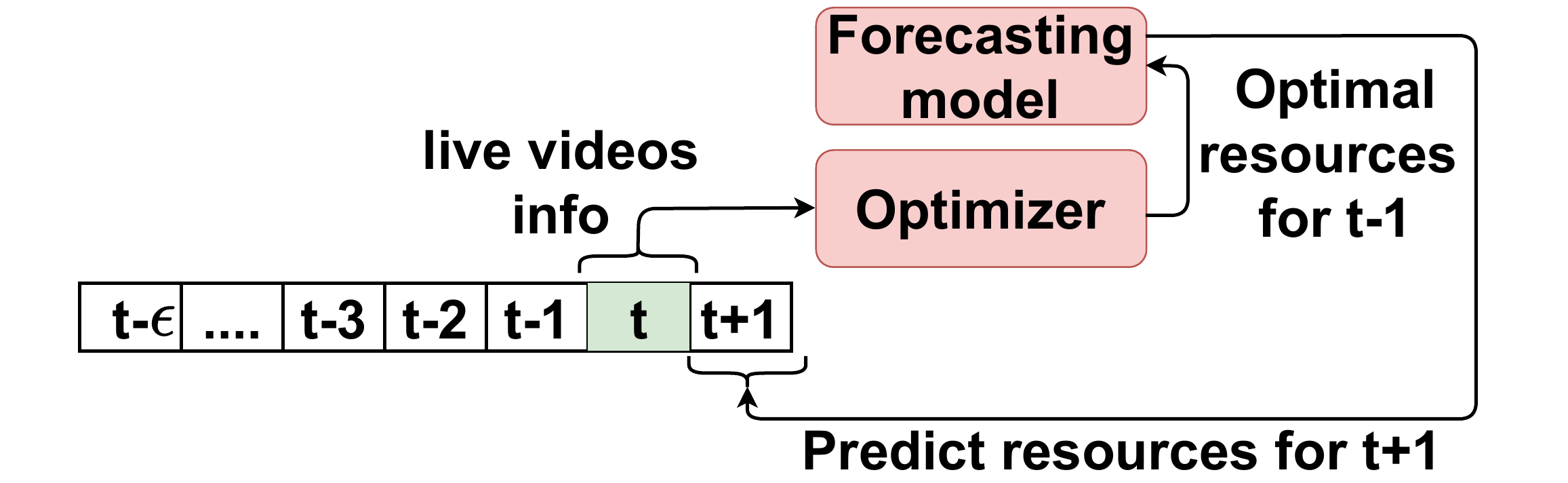}
  \caption{\small System timeline at the start of t: The offline optimizer receives the set of videos metadata of period t-1, to decide the optimal number of resources and update the forecasting models.} 
  \label{fig:timeline}
  \vspace{-0.2cm}
\end{figure}

Algorithm \ref{Proactive} presents the described proactive resource reservation and for better understanding we illustrate different steps of the algorithm in Fig. \ref{fig:timeline}.
\subsection{Online heuristic GNCA}
\begin{algorithm}[h]
\caption {Greedy Nearest Cheapest Algorithm (GNCA)}
\label{PGN}
\footnotesize
\begin{algorithmic}[1]
\State \footnotesize{\textbf{Input: $R$, \{$\zeta_1,...,\zeta_n$\},  \{$\mu_1,...,\mu_n$\}, $RTT$, $RI(t)$, $DI(t)$, $dissThreshold$, $V(t)$, $P(t)$, $B(t)$, $QB(t)$.
}}
\State \footnotesize{\textbf{Output:} $I(v, q^{tr},r^{tr})$, $W(v, q^{re},r^{tr}, r^{w})$
}
\For {$t \in{\{1,..,T\}}$}
\State \footnotesize{- Rank the videos in $V(t)$ in descending order based on popularity.}
\For {$video$ $v \in V(t)$ }
\State \footnotesize{- initialize current number of viewers CVN=0}
\State \footnotesize{- initialize dissatisfied number of viewers DVN=0}
\State \footnotesize{- initialize dissatisfaction percentage diss=0}
\State \footnotesize{- sortedR=R sorted in ascending order of prices of on demand}
\State \footnotesize{resources.}
\For {$region$ $r^{w} \in R$ }
\State \footnotesize{- Set allocated=0}
\State \footnotesize{- Set CVN=CVN+$p_{r^w}(q^r)$}
\State \footnotesize{- Sort the region in ascending order of distance from $r^{w}$.}
\State \footnotesize{- Check the requested quality $q^r$}
\For {$region$ $r^{tr} \in R$ }
 \If{$d_{r^{tr}r^{w}}$ $\leq \mathbb{D}$ \AND v with quality $q^r$ is not allocated \\ \quad \quad \qquad \qquad$\in r^{tr}$ \AND $RI(t) \in  r^{tr} > 0$}
 \State \footnotesize{- Transcode video to $q^r$ in $r^{tr}$ .}
 \State \footnotesize{- Set allocated=1.}
 \State \footnotesize{- $RI(t)=RI(t)-1$.}
  \EndIf
\EndFor
\If{allocated=0 and diss $\leq$ dissThreshold}
\For {$region$ $r^{tr} \in R$}
 \If {v with quality $q^r$ is not allocated $\in r^{tr}$ \\ \quad \quad \qquad \qquad \quad\AND $RI(t) \in  r^{tr} > 0$}
 \State \footnotesize{- Transcode video to $q^r$ in $r^{tr}$.}
 \State \footnotesize{- Set allocated=1.}
 \State \footnotesize{- Set DVN=DVN+$p_{r^w}(q^r)$.}
 \State \footnotesize{- Set diss=$DVN/CVN \times100$.}
 \State \footnotesize{- $RI(t)=RI(t)-1$.}
 \EndIf
 \EndFor
 \EndIf
\If{allocated=0}
\For {$region$ $r^{tr} \in sortedR$}
 \If{$d_{r^{tr}r^{w}}$ $\leq \mathbb{D}$ \AND v with quality $q^r$ is not allocated \\ \quad \quad \qquad \qquad \qquad$\in r^{tr}$ \AND $DI(t) \in r^{tr} > 0$}
 \State \footnotesize{- Transcode video to $q^r$ in $r^{tr}$.}
 \State \footnotesize{- Set allocated=1.}
 \State \footnotesize{- $DI(t)=DI(t)-1$.}
\EndIf
\EndFor
\EndIf
\EndFor
\EndFor
\EndFor
\end{algorithmic}
\end{algorithm}
When receiving the real-time incoming videos, the reserved resources forecasted by our prediction models should be used optimally to allocate the streams and serve viewers, with minimum costs and maximum QoS. For that, we present our GNCA heuristic, described in Algorithm \ref{PGN}. The inputs to the GNCA are the cost of renting on demand and reserved cloud instances at each cloud site, the RTT delay matrix, the hourly reserved instances at each cloud site, the hourly on demand cloud instances limit at each cloud site, the latency threshold and the dissatisfied viewers percentage threshold ($dissThreshold$) at each cloud site. The algorithm returns an allocation matrix that represents video representations needed at each cloud site, and a viewers serving matrix that represents from which cloud site viewers are served to meet the latency threshold and the requested quality. Let $RI(t)$ denote a set of reserved cloud instances at each cloud site for time period t based on the forecasting models prediction. $DI(t)$ is a set of on demand cloud instances limit at each cloud site for time step t. The algorithm starts by ranking the incoming videos in descending order based on popularity. In order to serve the geo-distributed viewers of each video, the algorithm starts first by serving the viewers using the reserved cloud instances. Next, it ranks the cloud sites $r^{tr}$ in $R$ in ascending order based on their RTT delay to the region of viewers $r^w$. The algorithm then traverses the list of ranked regions in order to find the nearest region that meets the delay threshold. If the video with that quality is not already transcoded and allocated in the nearest region, and if there are enough reserved resources in that region, the allocation is made, otherwise, the next cloud site is checked. For videos with low popularity, there is a probability that the number of reserved cloud instances that meet the latency threshold will not be sufficient. In this case, the next nearest cloud site will be used for allocation and some viewers will be dissatisfied by being served with a higher delay than the constraint.  The algorithm has a constraint on the percentage of viewers that can be dissatisfied and served with a higher latency. In fact, once the $dissThreshold$ is met, the algorithm starts renting on demand resources to serve the remaining viewers to meet the latency threshold. In this scenario, the algorithm ranks the cloud sites in ascending order based on the on demand cloud instances prices. The $dissThreshold$ constraint can be set based on the content provider' application requirements. For example, for popular applications, the content provider can sacrifice in terms of cost by wasting the remaining reserved instances, and decide not to dissatisfy users. On the other hand, for some applications, viewers of low popularity videos can be dissatisfied in order to exploit most of the reserved resources. It is worth mentioning, that GNCA serves the viewers from the nearest region when using the reserved instances because the renting cost in this case was already paid. However, when using the on demand resources the viewers are served from the cheapest region since the renting cost is paid on-the-fly. 
\begin{table}[h]
\captionsetup{font=footnotesize}
\footnotesize
\caption {\small Simulation parameters.}
\centering
\begin{tabular}{|l|l|}
\hline
\multicolumn{1}{|c|}{\textbf{Variable}}                                             & \multicolumn{1}{c|}{\textbf{Simulation Parameters}}                                                                                                                                    \\ \hline
\multicolumn{2}{|c|}{\textbf{Resource allocation optimizater/GNCA}}                                                                                                                                                                                                          \\ \hline
Video size for quality 240P                                                         & $\kappa(q_1)$= 0.405 Gbit                                                                                                                                                              \\ \hline
Video size for quality 360p                                                         & $\kappa(q_2)$= 0.495 Gbit                                                                                                                                                              \\ \hline
Video size for quality 480p                                                         & $\kappa(q_3)$= 0.603 Gbit                                                                                                                                                               \\ \hline
Video size for quality 720p                                                         & $\kappa(q_4)$= 0.738 Gbit                                                                                                                                                               \\ \hline
\begin{tabular}[c]{@{}l@{}}The computational cloud \\ instances prices\end{tabular} & \begin{tabular}[c]{@{}l@{}}$\zeta$ (cloud instance cost)\\  $\omega$ (migration cost)\\ $\eta$ (serving cost)\\ Amazon EC2 c5.large prices \cite{amazonDiscount}\end{tabular} \\ \hline
Delay threshold                                                                     & $\mathbb{D}$= 8.8ms, 120ms, and 180ms                                                                                                                                                   \\ \hline
\multicolumn{2}{|c|}{\textbf{GNCA}}                                                                                                                                                                                                                                          \\ \hline
 \begin{tabular}[c]{@{}l@{}}Percentage of dissatisfied viewers\end{tabular}                                                                     & $dissThreshold$= 0\%, 5\%, and 10\%                                                                                                                                                                     \\ \hline
\begin{tabular}[c]{@{}l@{}}Reserved cloud instances \\prices  \end{tabular}         &
\begin{tabular}[c]{@{}l@{}} $\mu$ (reserved cloud instance cost)\\ Amazon EC2 c5.large prices \cite{amazonDiscount}     \end{tabular}
                                   \\ \hline
\multicolumn{2}{|c|}{\textbf{Forecasting models}}                                                                                                                                                                                                                            \\ \hline
Neural models layers/neurons                                                        & One layer\textbackslash{}100 neurons                                                                                                                                                   \\ \hline
Activation function                                                                 & Rectified Linear Units (ReLU)                                                                                                                                                          \\ \hline
Optimization function                                                               & Adagrad                                                                                                                                                                                \\ \hline
Dropout                                                                             & 0.4                                                                                                                                                                                    \\ \hline
Minibatch                                                                           & 32                                                                                                                                                                                     \\ \hline
XGboost number of estimators                                                        & 1000                                                                                                                                                                                   \\ \hline
\end{tabular}
\label{param}
\end{table}
\begin{figure}[h]
\centering
 \includegraphics[scale=0.45]{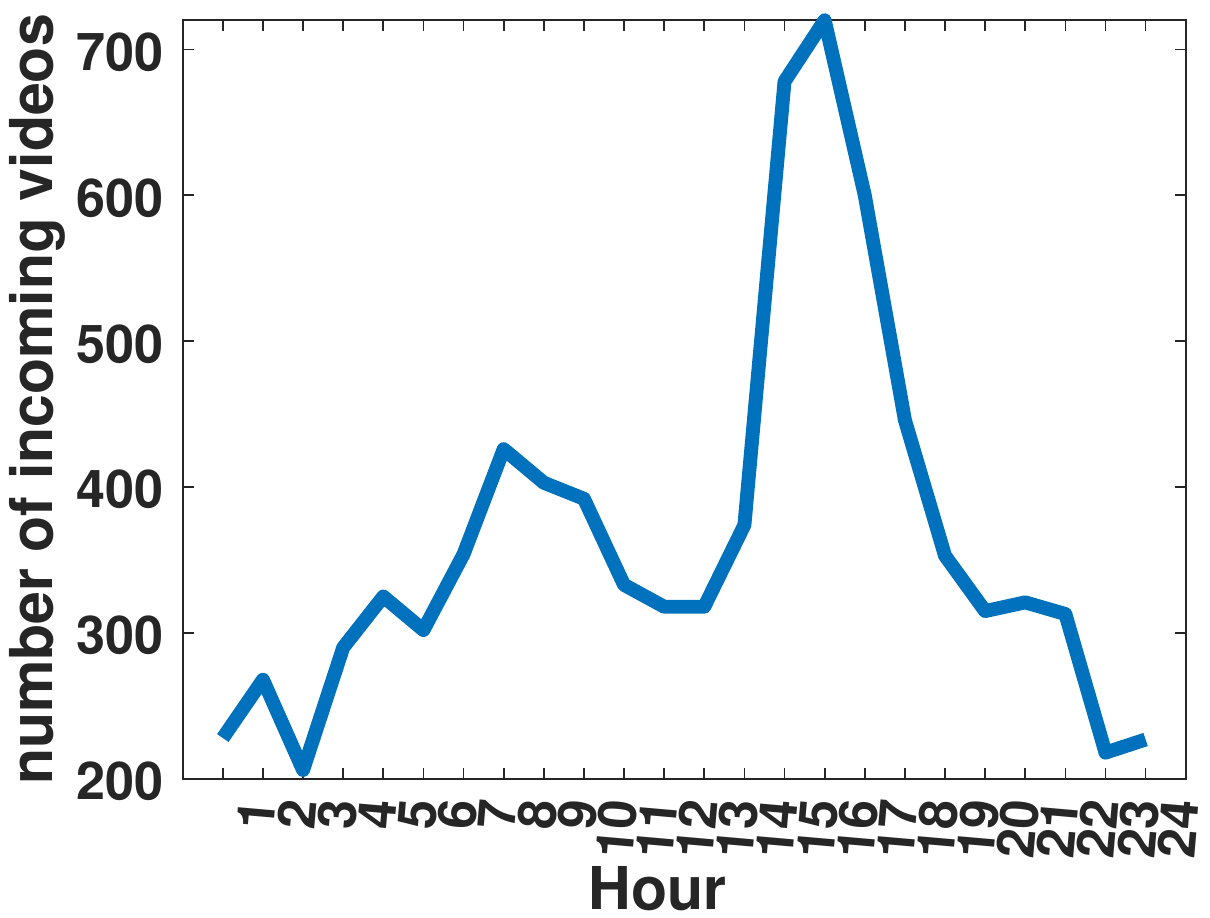}
  \caption{\small Number of hourly broadcasted videos in $23^{th}$ June 2018.}
  \label{fig:videosNum}
  \vspace{-0.4cm}
\end{figure}
\begin{table*}[h]
\captionsetup{font=footnotesize}
\footnotesize
\caption{ \small $R^{2}$ testing results: the optimizer is tested on 3 different latency thresholds and resulted in 3 datasets for each region. Five forecasting models were trained and the best model for each region is highlighted.}
\begin{adjustbox}{width=1\textwidth}
\footnotesize
\centering
\begin{tabular}{|c|c|c|c|c|c||c|c|c|c|c||c|c|c|c|c|}

\hline

\multirow{2}{*}{}&
\multicolumn{5}{c||}{\textbf{8.8ms dataset}}&
\multicolumn{5}{c||}{\textbf{120ms dataset}}&
\multicolumn{5}{c|}{\textbf{180ms dataset}}\\\cline{2-16} 
                & \textbf{GRU}  & \textbf{LSTM} & \textbf{MLP}  & \textbf{CNN} & \textbf{XGboost} & \textbf{GRU}  & \textbf{LSTM} & \textbf{MLP}  & \textbf{CNN} & \textbf{XGboost} & \textbf{GRU}  & \textbf{LSTM} & \textbf{MLP}  & \textbf{CNN} & \textbf{XGboost} \\ \hline
\textbf{Mumbai}     & 0.76          & 0.77          & \cellcolor{red!25}\textbf{0.81} & 0.66         & 0.76             & \cellcolor{red!25}\textbf{0.79} & 0.71          & 
0.70          & 0.67         & 0.58             & \cellcolor{red!25}\textbf{0.71} & 0.70          & 0.69          & 0.67         & 0.68             \\ \hline
\textbf{Seoul}      & 0.92          & 0.92          & \cellcolor{red!25}\textbf{0.93} & 0.86         & 0.92             & \cellcolor{red!25}\textbf{0.94} & \cellcolor{red!25}\textbf{0.94} & 0.92          & 0.89         & 0.93             & 0.89          & \cellcolor{red!25}\textbf{0.90} & \cellcolor{red!25}\textbf{0.90} & 0.87         & \cellcolor{red!25}\textbf{0.90}    \\ \hline
\textbf{Singapore}  & 0.92          & \cellcolor{red!25}\textbf{0.93} & 0.92          & 0.88         & \cellcolor{red!25}\textbf{0.93}    & \cellcolor{red!25}\textbf{0.95} & 0.94          & 0.94          &
 0.92         & 0.92             & \cellcolor{red!25}\textbf{0.94} & 0.93          & 0.93          & 0.93         & 0.91             \\ \hline
\textbf{China}      & \cellcolor{red!25}\textbf{0.94} & 0.93          & 0.93          & 0.91         & 0.93             & \cellcolor{red!25}\textbf{0.94} & \cellcolor{red!25}\textbf{0.94} & \cellcolor{red!25}\textbf{0.94} & 0.89         & 0.92             & 0.93          & \cellcolor{red!25}\textbf{0.94} & \cellcolor{red!25}\textbf{0.94} & 0.92         & 0.93             \\ \hline
\textbf{Frankfurt}  &\cellcolor{red!25} \textbf{0.85} & 0.83          & 0.82          & 0.74         & 0.76             & \cellcolor{red!25}\textbf{0.83} & 0.80          & \cellcolor{red!25}\textbf{0.83} & 0.81         & 0.67             & 0.78          & \cellcolor{red!25}\textbf{0.79} & 0.77          & 0.68         & 0.62             \\ \hline
\textbf{Paris}      & 0.78          & \cellcolor{red!25}\textbf{0.81} & 0.80          & 0.71         & 0.65             & \cellcolor{red!25}\textbf{0.73} & 0.72          & 0.73          & 0.61         & 0.72             & \cellcolor{red!25}\textbf{0.76} & 0.74          & \cellcolor{red!25}\textbf{0.76} & 0.75         & 0.75             \\ \hline
\textbf{Sao Paulo}  & 0.81          & \cellcolor{red!25}\textbf{0.82} & 0.80          & 0.72         & 0.82             & 0.78          & \cellcolor{red!25}\textbf{0.79} & 0.75          & 0.63         & 0.78             & 0.78          & 0.78          & 0.78          & 0.73         & \cellcolor{red!25}\textbf{0.81}    \\ \hline
\textbf{Ohio}       & 0.78          & 0.78          & 0.77          & 0.54         & 0.68             & \cellcolor{red!25}\textbf{0.77} & \cellcolor{red!25}\textbf{0.77} & 0.71          & 0.62         & 0.73             & \cellcolor{red!25}\textbf{0.71} & 0.70          & 0.67          & 0.51         & 0.64             \\ \hline
\textbf{Virginia}   & 0.78          & \cellcolor{red!25}\textbf{0.79} & 0.77          & 0.71         & 0.75             & 0.77          & 0.76          & \cellcolor{red!25}\textbf{0.80} & 0.58         & 0.78             & \cellcolor{red!25}\textbf{0.71} & 0.61          & 0.55          & 0.47         & 0.52             \\ \hline
\textbf{California} & 0.68          & \cellcolor{red!25}\textbf{0.73} & 0.66          & 0.62         & 0.60             & 0.73          & 0.73          & \cellcolor{red!25}\textbf{0.78} & 0.69         & 0.64             & \cellcolor{red!25}\textbf{0.81} & 0.80          & 0.77          & 0.62         & 0.70             \\ \hline
\end{tabular}
\end{adjustbox}
\label{table:results}
\vspace{-0.4cm}
\end{table*}
\section{Performance Evaluation}\label{section:performance}
\subsection{Simulation settings}
\subsubsection{Resource allocation optimizer}
 As described in section \ref{section:Dataset}, we used our collected Facebook videos metadata \cite{dataset} as an input to the optimizer. Also, we opted for AWS as a geo-distributed cloud platform, and we chose to adopt $n=10$ AWS cloud sites. We will show the performance of our system on a period $T=24$ hours corresponding to $23^{rd}$ June 2018. The optimization was run on hourly-based videos' arrival; $t=1$ hour (see section \ref{section:ProblemFormulation}). More specifically, Fig. \ref{fig:videosNum} presents the number of hourly broadcasted videos, on which we will test the performance of our optimal formulation. For simplicity, we assume that all videos have the same length equal to 1 hour as renting Amazon cloud instances is processed on hourly basis. Furthermore, we choose, in this work, to serve all viewers with the requested bitrates, which means with the higher satisfaction level in terms of quality. The size of a content depends on its bitrate, as illustrated in Table \ref{param}. Next, following the parameters adopted in \cite{bilal2018qoe,BACCOUR2020982}, the round trip time (RTT) matrix to find $d_{r^{tr}r^{w}}$ is created by examining the ping times between different geo-distributed cloud sites \cite{wondernetwork}. Also,  we follow the charging model of Amazon EC2 c5.large  to calculate $\zeta$, $\omega$ and $\eta$ \cite{amazonDiscount}. Finally, we consider different latency thresholds $\mathbb{D}$ to evaluate the performance of our system, specifically 8.8 ms, 120 ms and 180 ms. Note that the minimum serving latency, incurred when delivering the stream from the closer data center is equal to 8.8 ms \cite{bilal2018qoe}.
 \begin{figure*}[h]
  \centering
\mbox{
  \subfigure[]{\includegraphics[scale=0.32]{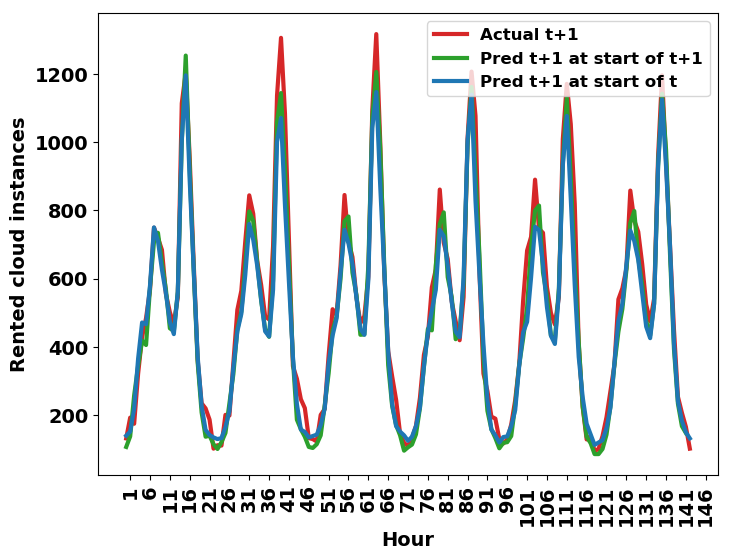}}
  \subfigure[]{\includegraphics[scale=0.32]{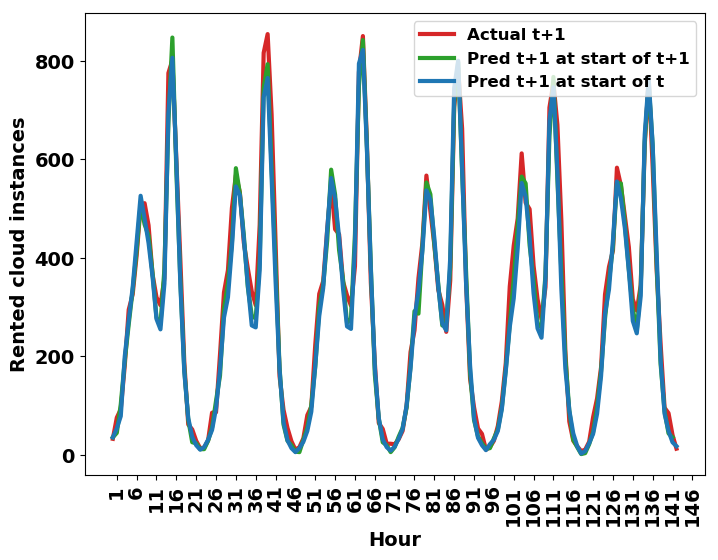}}
  \subfigure[]{\includegraphics[scale=0.32]{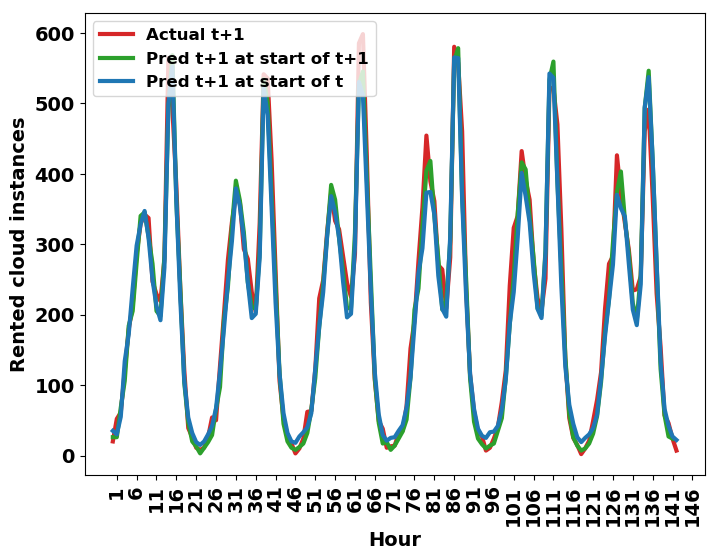}}
    }
  \caption{\small Hourly actual vs predicted cloud instances at Singapore datacenter: (a) 8.8ms latency threshold, (b) 120ms latency threshold, (c) 180ms latency threshold.}
  \label{fig:predVsActual_Singapore}
\end{figure*}
\subsubsection{Forecasting models} To create the geo-distributed datasets, we run our optimizer on the metadata of videos collected between $3^{rd}$ to $30^{th}$ June 2018, while using  different thresholds. As the output of the optimizer is the number of distributed resources to rent, we divided the data and created three datasets for each cloud region containing the required instances in this site at each slot of time ($t$=1 hour), for different thresholds. Finally, these time series data are reconstructed into a supervised learning, where the window size is set to  $\varepsilon $=24. Meaning, at the beginning of each time frame $t$, the previous 24 hours decisions will be fed to different models, in order to predict the required resources for the next slot $t+1$, as depicted in Fig. \ref{fig:timeline}. We note that we fixed $t$ to 1 hour and we considered hourly resource prediction, because we are constrained by the size of our collected dataset \cite{dataset} that has only few months data. If we dispose of a larger dataset, we can divide the time series into $t$ equal to days or weeks. In this way, resource forecasting and renting can be performed earlier and with lower costs. Finally, the data is split into a training part from the $3^{rd}$ to the $24^{th}$ of June, and a testing part from the $25^{th}$ to the $30^{th}$ of same month. The learning tasks are implemented using Scikit-learn Python library \footnote{https://scikit-learn.org/stable/index.html}.

In our experiments, we adopted a gridsearch to find the best hyperparameters  that show the higher accuracies. These parameters are summarized in Table \ref{param}.
\subsubsection{GNCA}
We run our heuristic using the three mentioned latency thresholds. As our heuristic uses the reserved transcoding resources to allocate real-time incoming  videos, we used the predicted transcoding resources for the 24 hours of $25^{th}$ June 2018, as an input to the GNCA. Since our algorithm considers using on demand resources in case the reserved are not sufficient, and due to the fact that the number of on demand resources supply per region is limited, we set the hourly regional on demand cloud instances limit to 500.

As aforementioned, in case the reserved instances are not enough to meet the delay constraint, the system can use the remaining reserved instances and dissatisfy a certain percentage of viewers by serving them with a higher latency. In order to study the effect of dissatisfying the viewers on the latency and the cost, we varied the tolerated dissatisfaction to 0\% and 10\%. Finally, the prices of the reserved cloud instances are obtained from the Amazon pricing website \cite{amazonDiscount}.
\subsection{Simulation results} 
\subsubsection{Optimal transcoding resources for historical videos}
\begin{figure}[h]
\centering
	\mbox{
	    \hspace{-6.5mm} \subfigure[\label{fig:optimalCost}]{\includegraphics[scale=0.33]{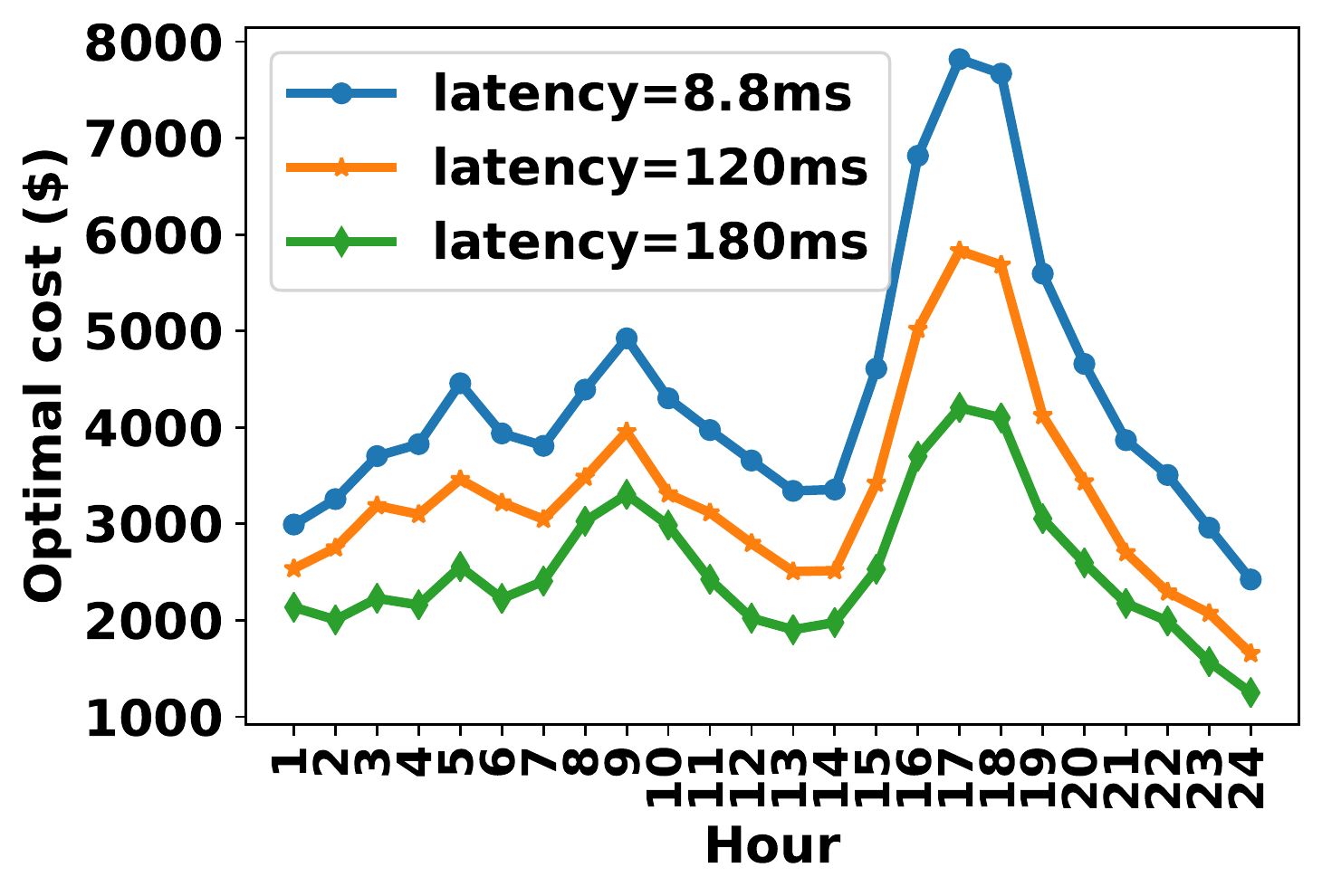}
	   }
	     \hspace{-4.5mm}
	     \subfigure[\label{fig:AvrLatency}]{\includegraphics[scale=0.33]{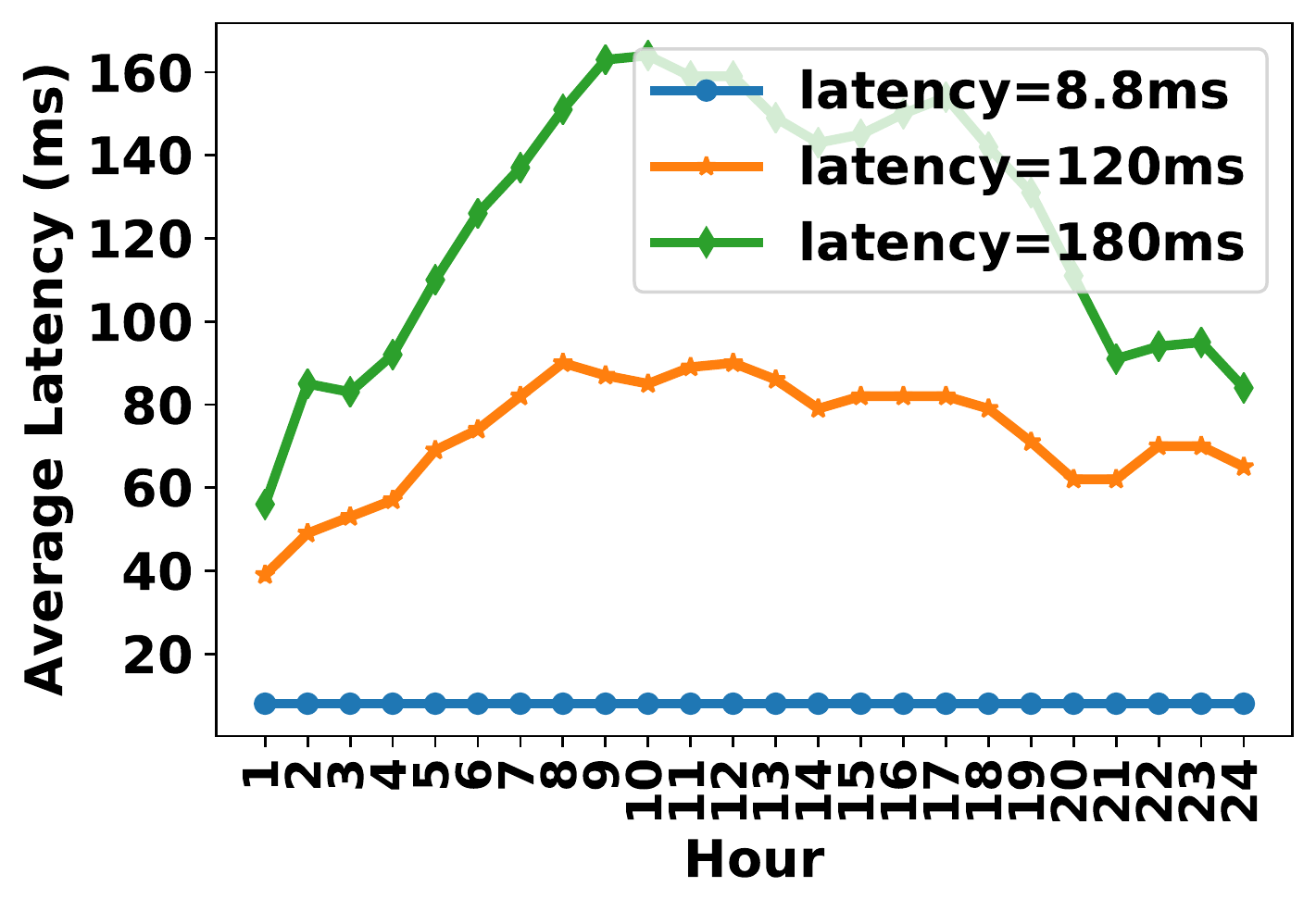}}
} 
	\caption{\small Simulation with historical videos of $23^{rd}$ June 2018: (a) Hourly optimal cost, (b) Hourly average latency. }
	\label{fig:opt}
	\vspace{-0.3cm}
\end{figure}
To evaluate the performance of the optimizer and analyse the resource allocation strategy, we examined two metrics which are the cost and the average latency. We recall that this cost includes the rental, migration and serving prices and that the average delay is defined as the average round-trip delay to serve all viewers of all videos. Fig. \ref{fig:optimalCost} and Fig. \ref{fig:AvrLatency} illustrate, respectively, the hourly overall cost and the average latency of the optimization results. Fig. \ref{fig:optimalCost}
shows that a trade-off between the video serving delay and the network cost can be established. Specifically, when the required delivery delay is very low (8.8 ms), the optimal cost is very high as the crowdsourcing system is forced to host, transcode and serve the viewers from their adjacent site. Relaxing the latency constraint contributes to reducing the cost, as the system can allocate the content in a lower cost site, while achieving higher delays. Hence, to maximize the viewers' satisfaction, the content provider can sacrifice of cost or vise versa, depending on the crowdsourcing platform requirements. We note that the fluctuating pattern of the hourly cost is justified by the fact that videos arrival is varying over time and some hours present peaks compared to valley periods (see Fig. \ref{fig:videosNum}). Fig. \ref{fig:AvrLatency} depicts the average delay achieved by the system hourly. We can see that, when fixing the threshold to 8.8 ms, the system always sticks to this delivery delay, as it presents the latency to serve the viewers from their closer site. Allocating the content in other sites means non-respect of the delay constraint. When fixing higher thresholds, lower latencies are achieved, as the system managed to find the optimal cost with low delivery delays; e.g., maximum latency of 100 ms if the threshold is fixed to 120 ms and 170 ms if it is fixed to 180 ms. 
\begin{figure*}[h]
  \centering
    \mbox{
  \subfigure[]{\includegraphics[scale=0.37]{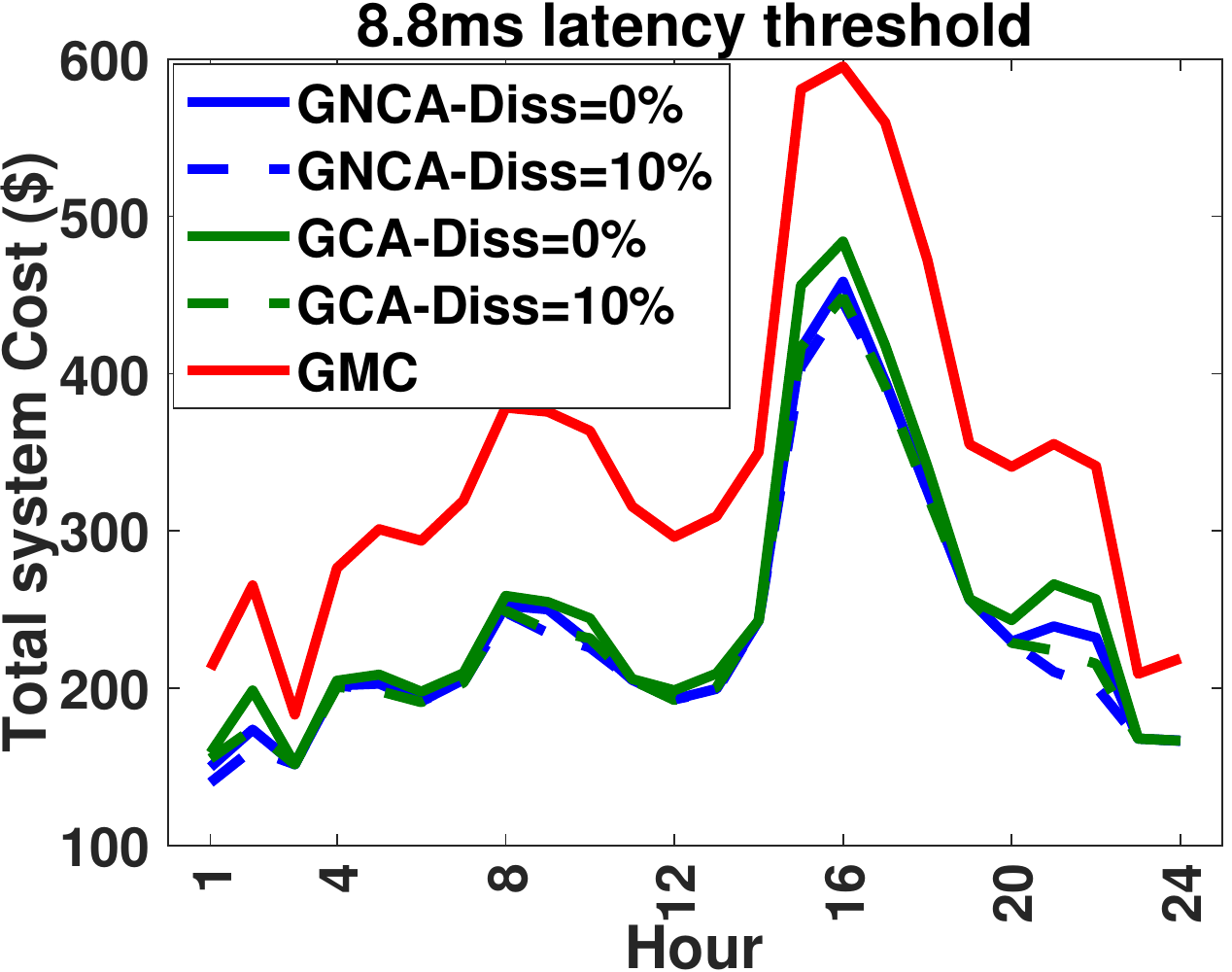}}
  \subfigure[]{\includegraphics[scale=0.37]{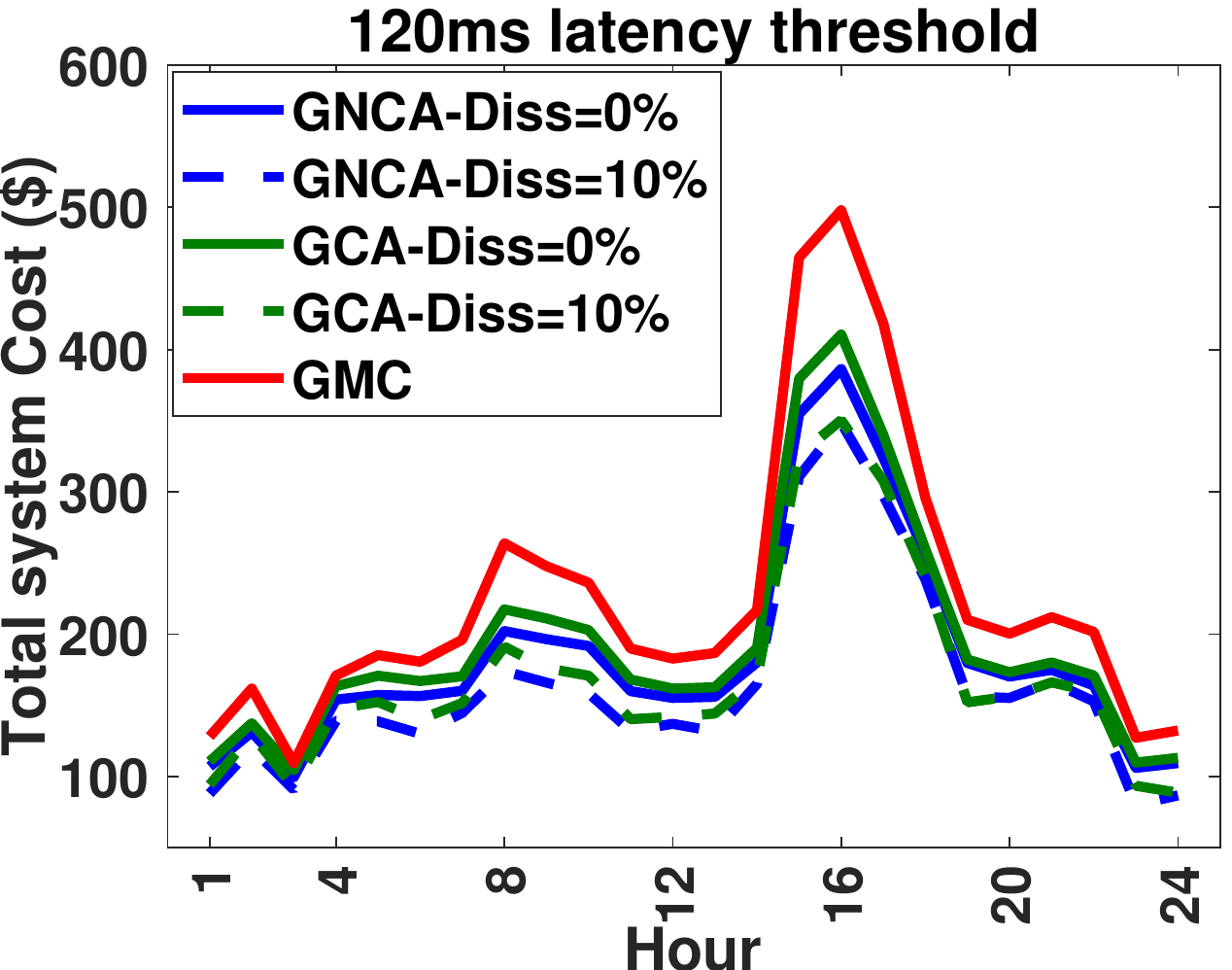}}
  \subfigure[]{\includegraphics[scale=0.37]{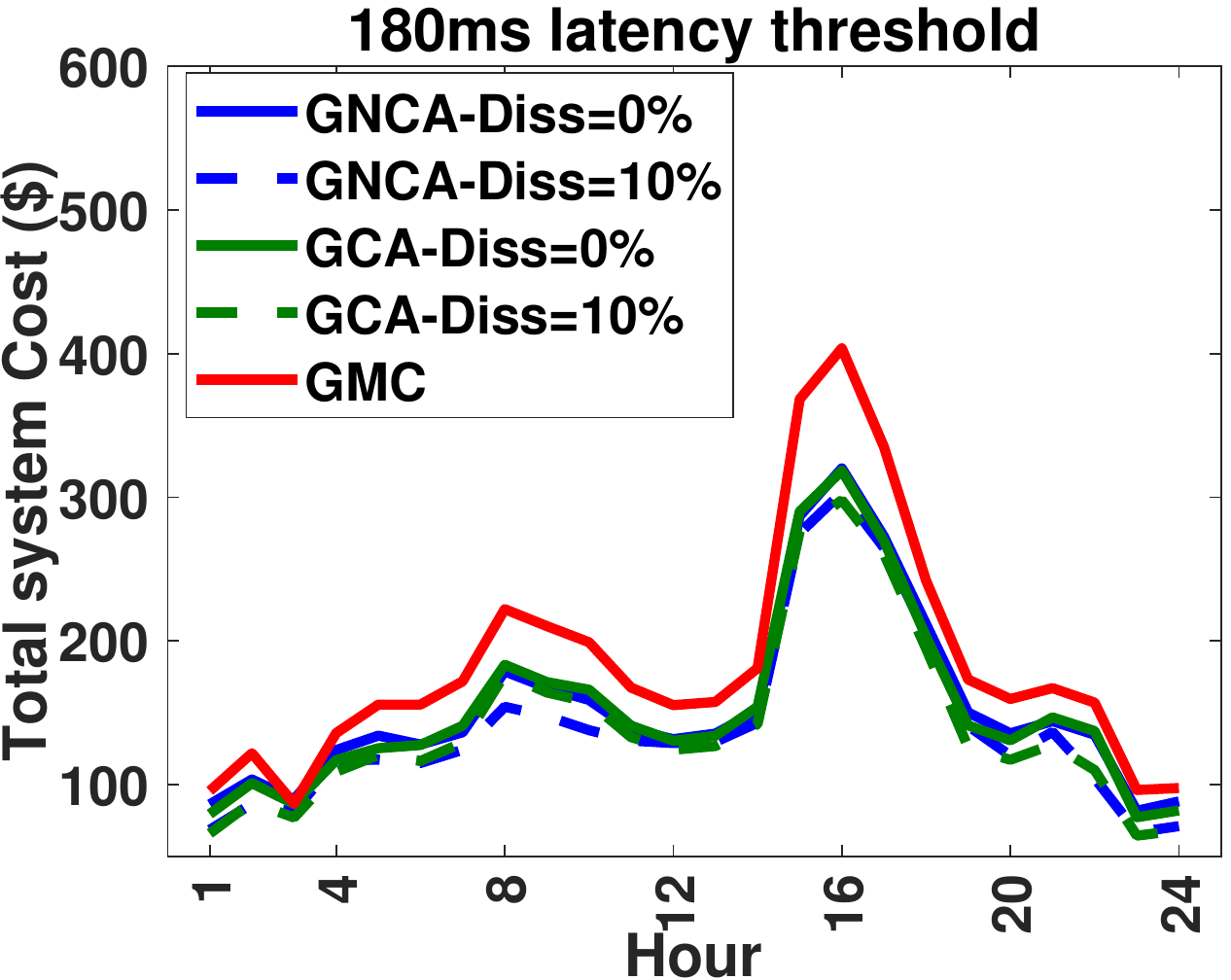}}
    }\\
            \mbox{
  \subfigure[]{\includegraphics[scale=0.37]{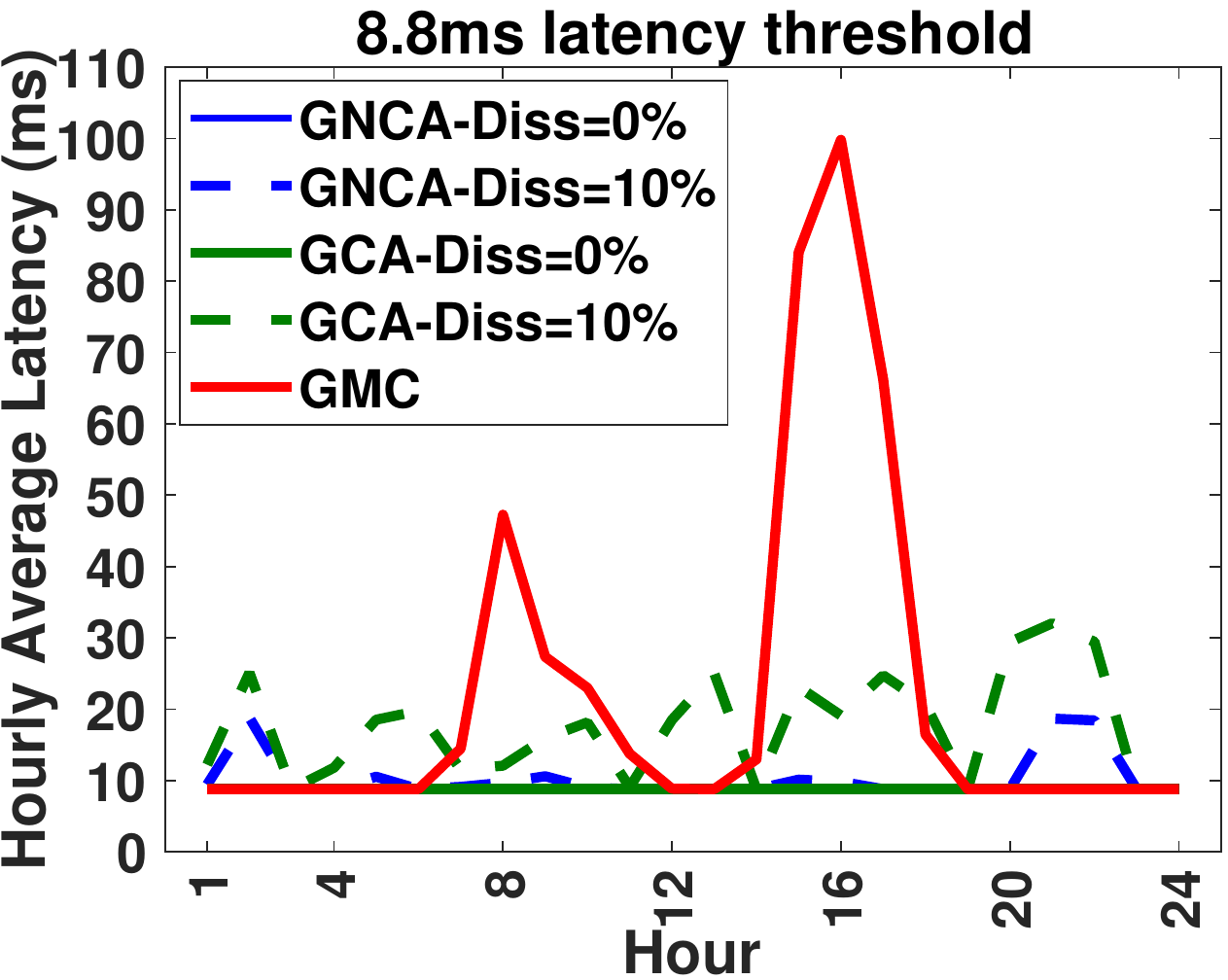}}
  \subfigure[]{\includegraphics[scale=0.37]{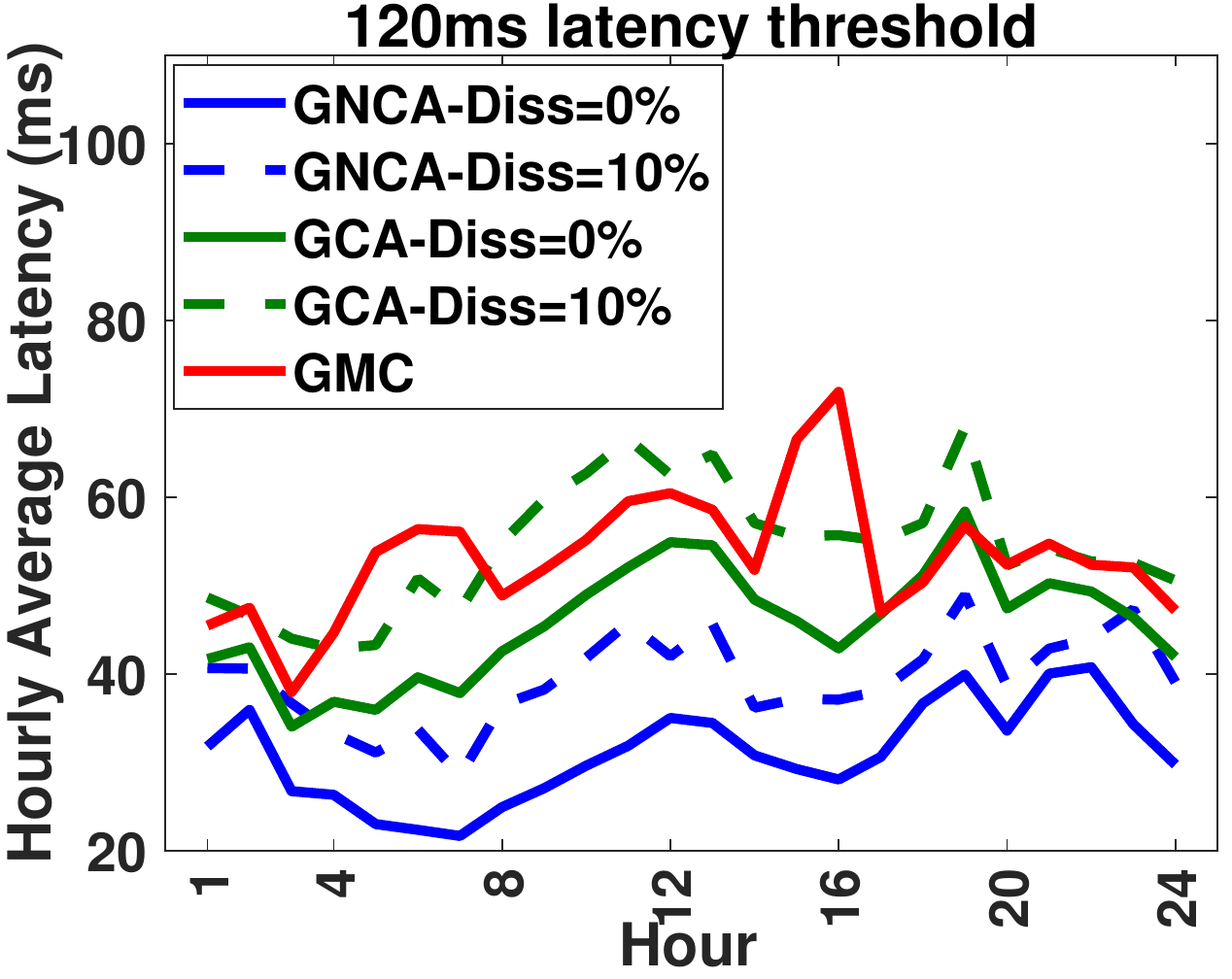}}
  \subfigure[]{\includegraphics[scale=0.37]{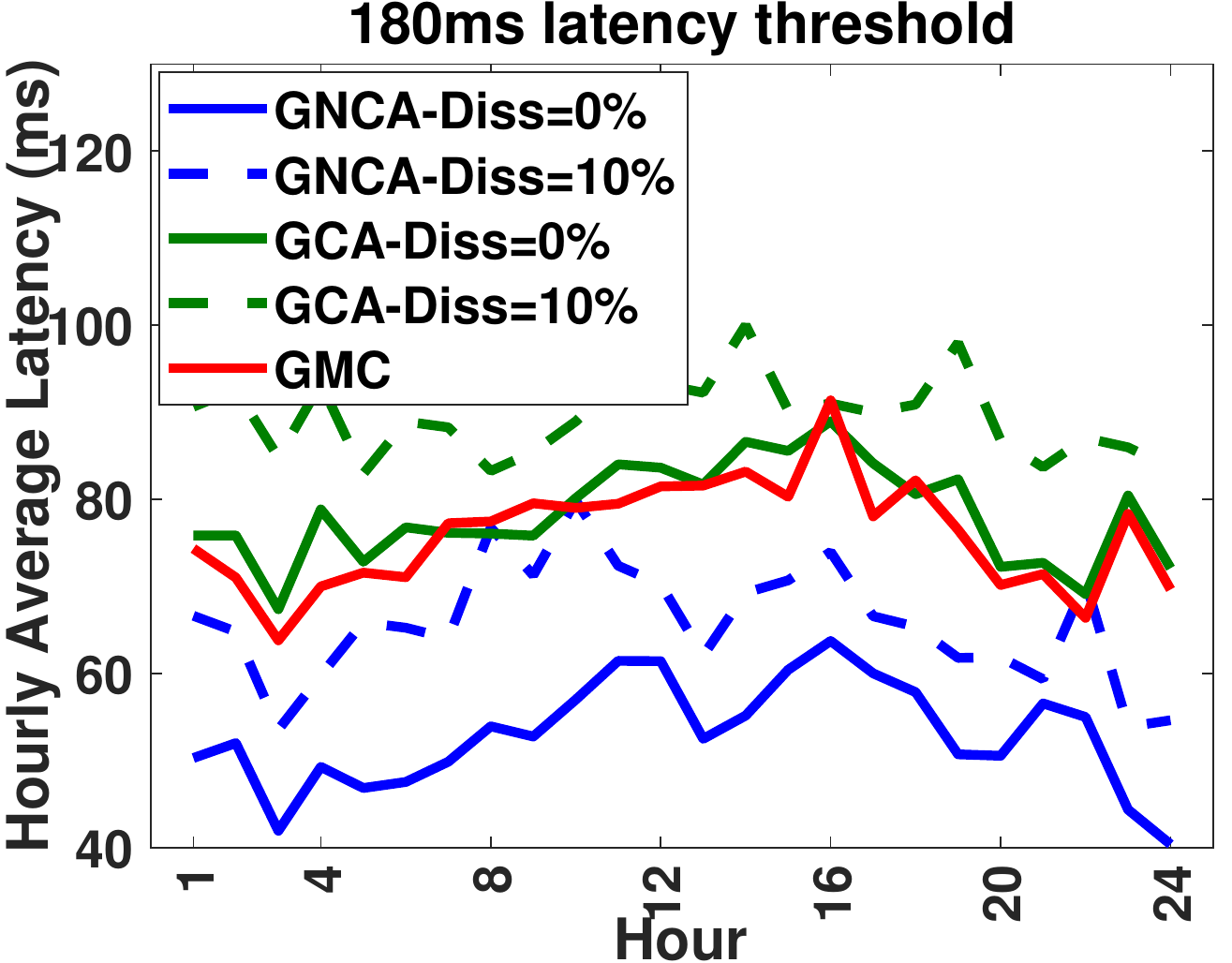}}
    }
  \caption{\small (a), (b), (c): Total system cost of GCA vs GMC vs GNCA with 2 dissatisfaction thresholds. \\(d), (e), (f): Hourly average latency of GCA vs GMC vs GNCA with 2 dissatisfaction thresholds.}
  \label{fig:totalCost_GNCAvsGCAvsGMC}
\end{figure*}
\subsubsection{The performance of the forecasting models}
As previously described, the results of the optimization serve to create our geo-distributed datasets. These datasets are trained using multiple machine learning techniques to choose the best one that predicts the required resources for the upcoming time slot. Table \ref{table:results} illustrates the testing results. We can see that LSTM, GRU, and MLP presented better performance for most of the regions. Meanwhile, XGboost achieved good results for some cloud sites and CNN showed the worst performance among all models.
These results were expected as LSTM and GRU are efficient in manipulating sequential data and memory state.  MLP can also perform as good as RNN family, if we model time connections properly by using as an input the $\varepsilon$ last time steps and predict the $(\varepsilon+1)^{th}$ time step, as we did in our work. Also, if the time series are short ($\varepsilon$ =24), the relationship through time will not be complex. In this case, a regular MLP can give good accuracy. XGboost is known by its performance compared to traditional machine learning techniques owing to its design, which enables it to achieve good results. Finally, CNN is mainly efficient for spatial data, such as images, which justifies its low performance in our architecture.

The performance of different ML algorithms varies from a region to another and there is no model that performs the best for all. Specifically, as we can see in the table, the prediction accuracy is higher in some regions, including China, Singapore, and Seoul, compared to other sites having lower accuracy such as California and Virginia. This can be explained by the fact that the distribution of viewers is different among cloud sites. This distribution results in distinct data pattern in each cloud site. By dealing with different patterns, ML models showed different accuracies, depending on the region. However, as we are adopting a distributed system and independent cloud sites, it is acceptable to  have different datasets and different forecasting models in each cloud site. We, also, noticed that, when varying the delay threshold, the performance of the model changes for the same cloud region. This can be explained by the fact that changing the delay constraints results in different optimizer decisions, hence, different dataset shape. Fig. \ref{fig:predVsActual_Singapore} presents the predicted number of cloud instances compared to the optimal number required to minimize the cost and maximize the QoS, during the period between $25^{th}$ to $30^{th}$ June 2018 (144 hrs) at Singapore cloud site. We can see that our prediction results achieve close performance compared to the optimal allocation. Furthermore, Predicting the amount of cloud instances for period t+1 at the start of the period t is our proposed forecasting model results, while predicting t+1 at the start of t+1 is the online and complex scenario. We observed that in some hours the online prediction slightly outperformed the offline prediction because as we predict ahead of time, the predictions will degrade. 
\begin{figure*}[h]
  \centering
               \mbox{
  \subfigure[]{\includegraphics[scale=0.371]{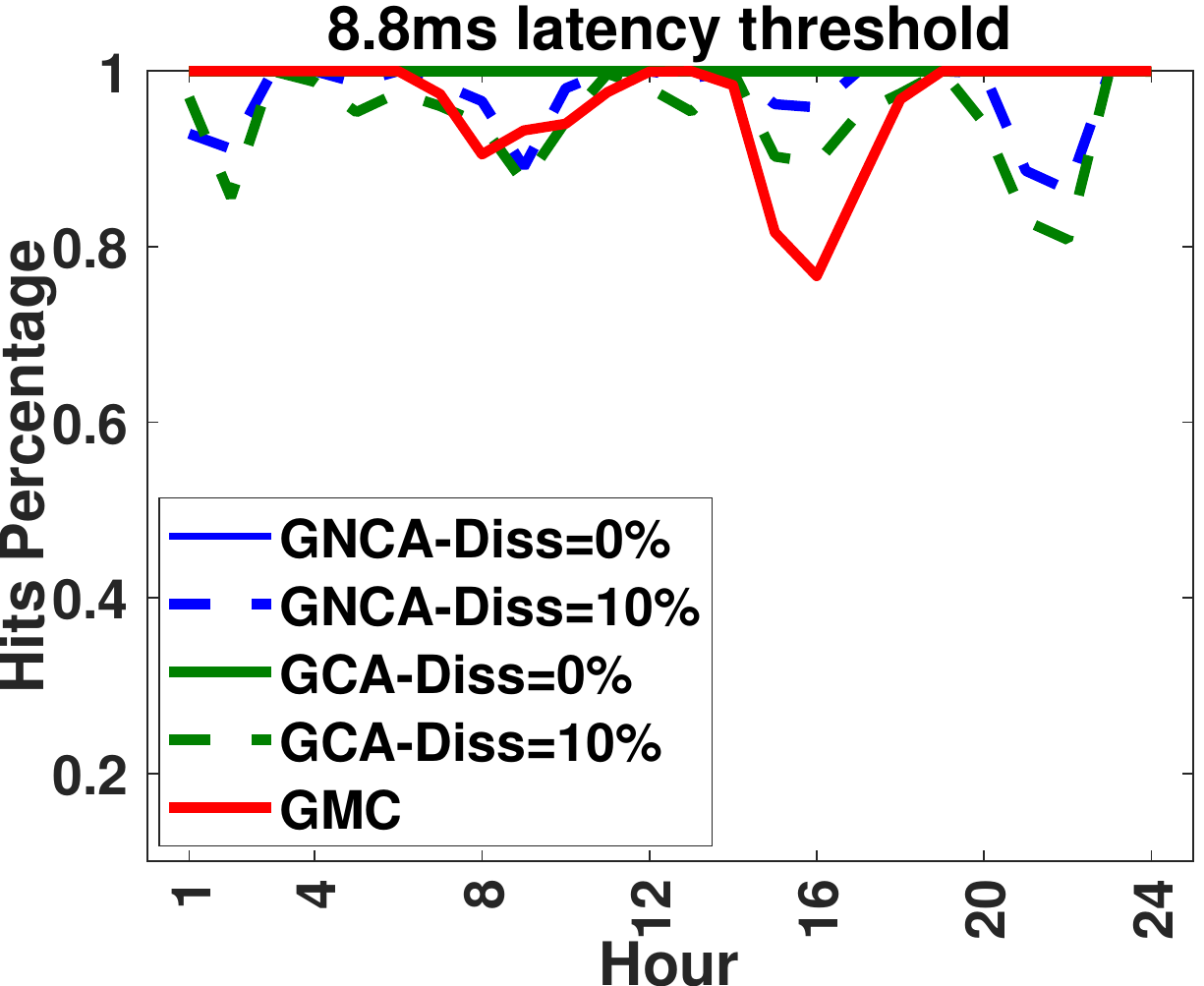}}
  \subfigure[]{\includegraphics[scale=0.37]{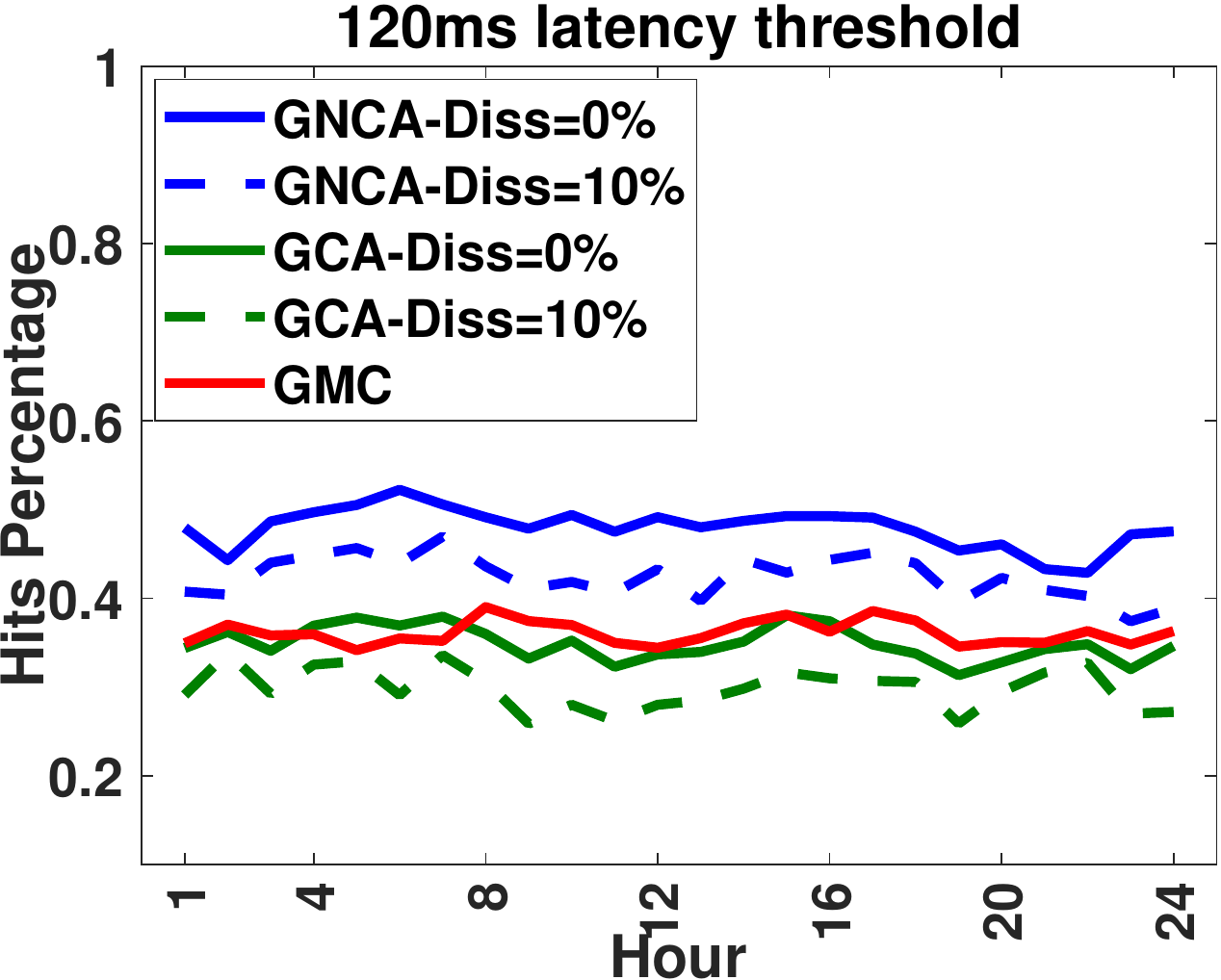}}
  \subfigure[]{\includegraphics[scale=0.37]{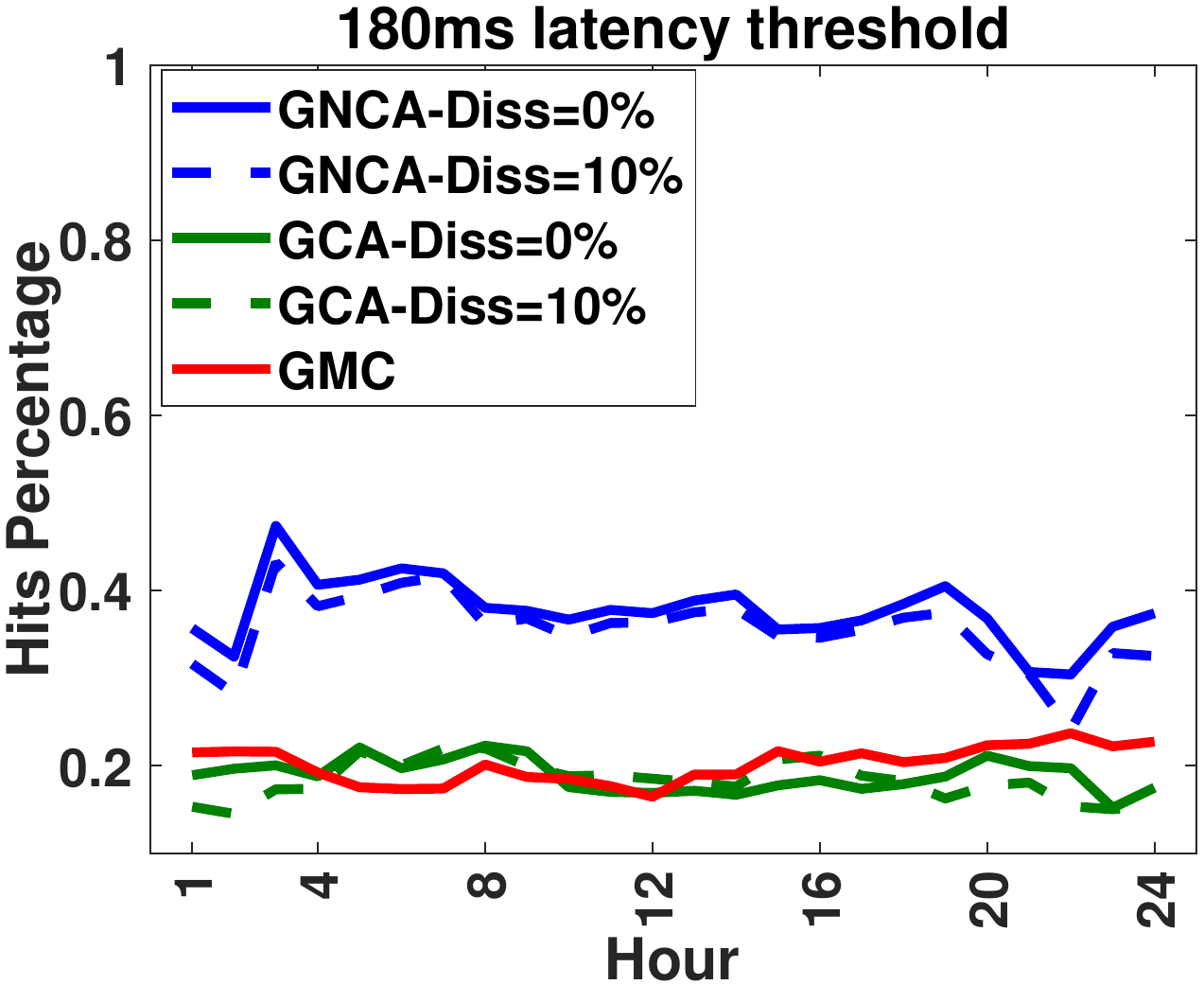}}
    }\\
               \mbox{
  \subfigure[]{\includegraphics[scale=0.37]{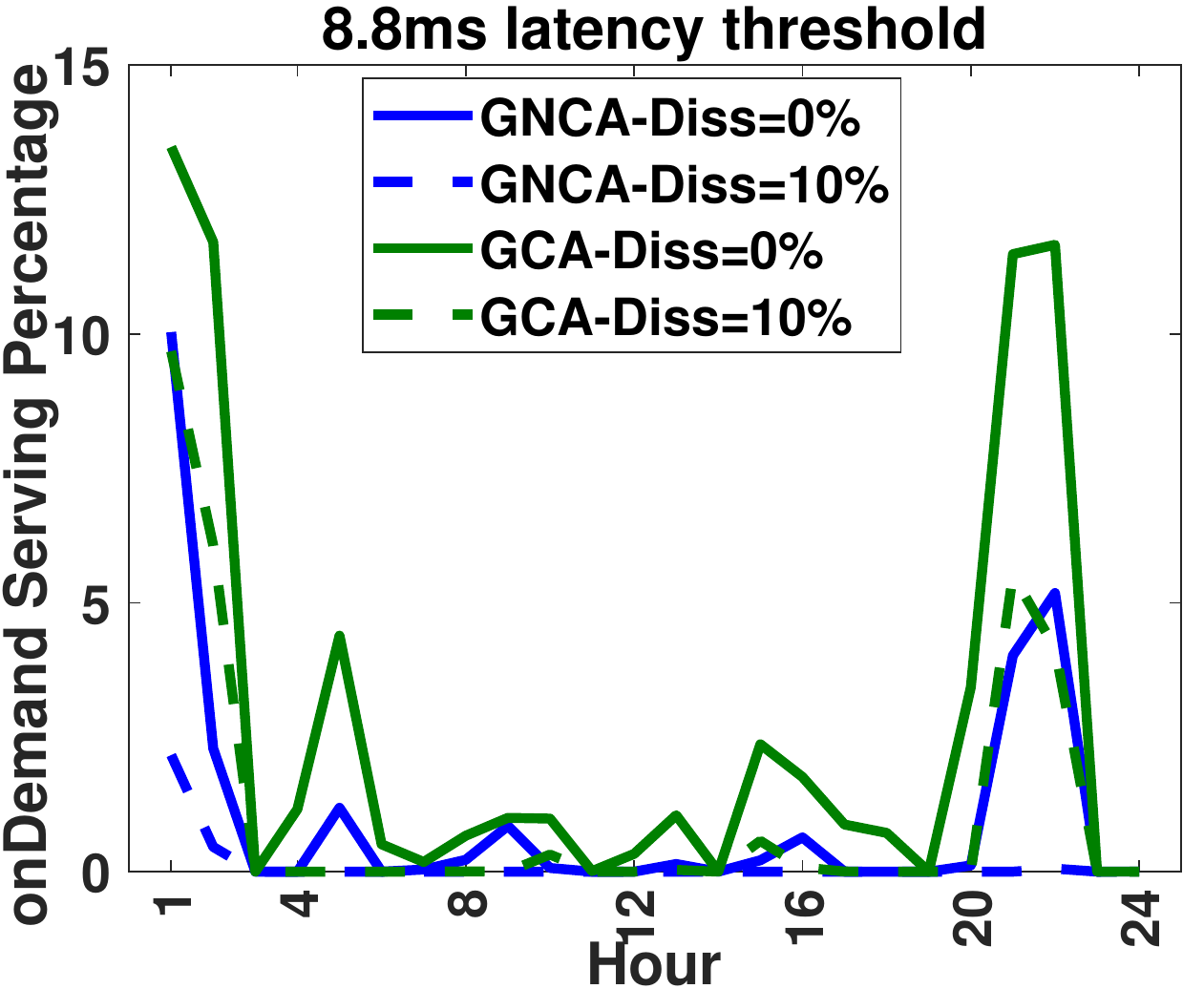}}
  \subfigure[]{\includegraphics[scale=0.37]{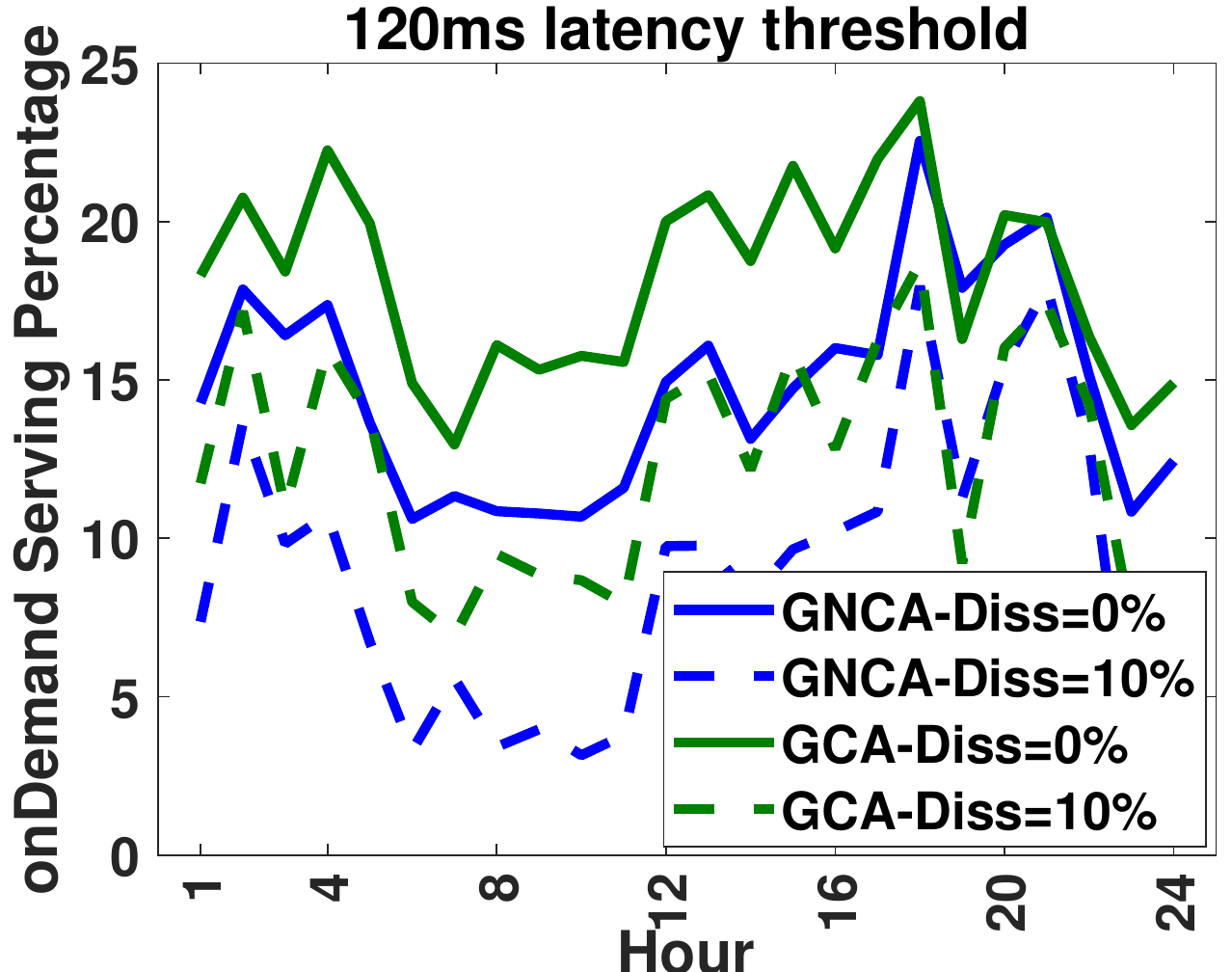}}
  \subfigure[]{\includegraphics[scale=0.37]{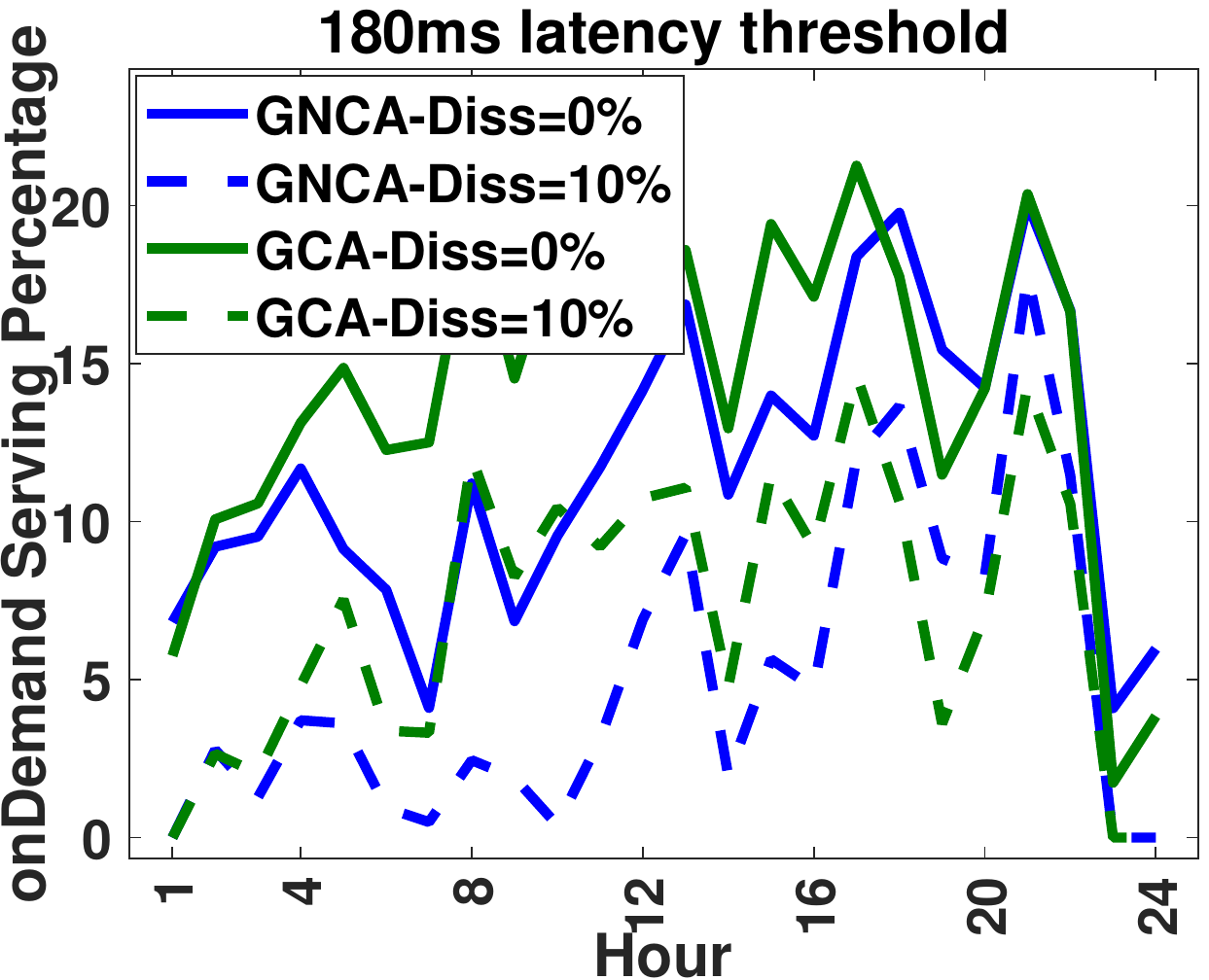}}
    }
  \caption{\small (a), (b), (c): Hit percentage of GCA vs GMC vs GNCA with 2 dissatisfaction thresholds.
  \\ (d), (e), (f): On demand serving percentage pf GNCA vs GCA. GMC always uses on demand strategy.}
  \vspace{-0.3cm}
  \label{fig:onDemand_PGNvsGMC}
\end{figure*}
\subsubsection{The performance of the GNCA greedy heuristic} Fig. \ref{fig:totalCost_GNCAvsGCAvsGMC}a, \ref{fig:totalCost_GNCAvsGCAvsGMC}b, \ref{fig:totalCost_GNCAvsGCAvsGMC}c show the total system cost of our algorithm against the greedy minimal cost (GMC) algorithm \cite{bilal2018qoe}, with latency threshold of 8.8ms, 120ms, and 180ms respectively. GMC  considers solely on demand resource allocation and chooses always the cheapest site that respects the delay threshold. The results show that GNCA achieves a significant decrease in the hourly system cost, due to the proactive reservation with reduced prices. Fig. \ref{fig:totalCost_GNCAvsGCAvsGMC}b and Fig. \ref{fig:totalCost_GNCAvsGCAvsGMC}c show that increasing the percentage of tolerated dissatisfaction decreases the cost because less on demand resources are required. Note that 0\% of dissatisfaction means that all viewers are served while respecting the latency threshold, and 10\%  of tolerated dissatisfaction means that no more than 10\% of viewers are  served with a latency higher than the threshold. We can, also, observe that the difference in system cost between GMC and GNCA is higher when the latency threshold is set to 8.8ms because the GMC system is forced to serve the viewers from their regions which requires a high number of on demand resources and leads to a big increase in cost. We can conclude that there is a trade-off between the total system cost and viewers dissatisfaction, and that the content provider can sacrifice in terms of cost and satisfy all viewers by serving them with the required latency or vice versa according to the application requirements. It is worth noting in Fig. \ref{fig:totalCost_GNCAvsGCAvsGMC}a that the total system cost was only slightly affected on some hours such as hour 1, 2, 21, and 22 because the reserved instances were sufficient in most of the hours. This is proven in Fig. \ref{fig:onDemand_PGNvsGMC}d, which presents the hourly percentage of viewers served with on demand instances, it is clear that even when the dissatisfaction threshold was set to 10\% the percentage of viewers served with on demand resources is 0\% in most of the hours, while we can notice that in hours 21 and 22 for example the percentage is higher. 

Our proposed GNCA serves the viewers using the reserved instances from their nearest region as the cost was paid anyway during reservation. However, the serving is done from the cheapest region in case on demand resources were used as the cost is paid on-the-fly. Besides comparing our algorithm against GMC, we implemented another baseline, Greedy Cheapest Algorithm (GCA). GCA mimics our approach in reserving cloud instances proactively, however, it serves the viewers from the cheapest region in all scenarios whether resources are reserved or on demand.  As presented in Fig. \ref{fig:totalCost_GNCAvsGCAvsGMC}a, Fig. \ref{fig:totalCost_GNCAvsGCAvsGMC}b, and Fig. \ref{fig:totalCost_GNCAvsGCAvsGMC}c, we observe that GNCA achieves a lower cost than GCA, and that GMC is always higher in cost than GNCA and GCA. Moreover, as presented in Fig. \ref{fig:totalCost_GNCAvsGCAvsGMC}d, Fig. \ref{fig:totalCost_GNCAvsGCAvsGMC}e, and Fig. \ref{fig:totalCost_GNCAvsGCAvsGMC}f, we notice that GNCA achieves the lowest hourly average latency all the time. We note that when the latency threshold was set to 8.8ms as shown in Fig. \ref{fig:totalCost_GNCAvsGCAvsGMC}d, the perceived average latency is always 8.8 ms for GNCA and GCA, when the $dissThreshold$ is equal to 0\%. We also observe that GMC could not achieve an average hourly latency of 8.8ms on some hours because the regional on demand instances were insufficient and some viewers were dissatisfied and served from other regions not meeting the latency constraint.

We also evaluated our heuristic in terms of hit percentage. The hit rate is defined as the percentage of viewers that are served from their closer data center. As the number of viewers that receive the live video from their adjacent/closer cloud region increases, the hit rate increases accordingly. This metric helps to identify the percentage of users served with minimum delays. We notice that GNCA outperforms both GMC and GCA as shown in Fig. \ref{fig:onDemand_PGNvsGMC}b and Fig. \ref{fig:onDemand_PGNvsGMC}c. This can be attributed to the fact that GNCA is serving the viewers from their nearest regions unless the reserved instances are not sufficient or the system is obliged not to dissatisfy the viewers. However, GMC and GCA are allocating resources on the cheapest sites, which reduces the achieved hit ratio. When setting the latency threshold to 8.8ms as presented in Fig. \ref{fig:onDemand_PGNvsGMC}a, it can be observed that both GNCA and GCA achieved a hit percentage of 100\% when the dissatisfaction threshold was set to 0\% because the system is forced to serve the viewers from their region. GMC, on the other hand, was not able to achieve hit percentages of 100\% in all hours because it was obliged to dissatisfy viewers due to the constraint on regional cloud instances.

As discussed, GNCA outperformed both GMC and GCA in terms of total system cost, average latency, and hit percentage. This is attributed to the fact that GNCA is using a lower number of on demand resources and depends more on the reserved cloud instances which is up to 75\% less in price. Fig. \ref{fig:onDemand_PGNvsGMC}d, Fig. \ref{fig:onDemand_PGNvsGMC}e, and Fig. \ref{fig:onDemand_PGNvsGMC}f present the percentage of viewers served with on demand resources. On demand resources are defined as the instances that are allocated and payed on-the-fly, in case all reserved resources are exhausted. Paying and serving on-the-fly result in additional costs and larger serving delays caused by booting the servers. This metric helps to identify the efficiency and the performance of the geo-distributed resource allocation algorithm in using the pre-reserved resources and in reducing the need to rent on-demand servers. Hence, when the percentage of on-demand serving is higher, it means higher costs are incurred and it shows that the resource allocation algorithm is less efficient. We can observe that GNCA achieved less percentage than GCA when setting the dissatisfaction threshold to 0\% and 10\%, which means that our allocation strategy outperforms GCA in exploiting the reserved servers. We remind that GMC serves 100\% of the viewers with on-demand resources.

\subsubsection{Discussion}
Bilal et al. \cite{bilal2018qoe} assumed, in their work (GMC), that all viewers watch each video from one region. 
For a fair comparison, we redeveloped the GMC algorithm according to our simulation settings, in order to use our dataset where viewers are located in multiple regions. Given that $V$ is the number of incoming videos and $R$ is the number of regions, the time complexity of GMC is O$(VR^2)$, which is the same complexity as our online GNCA. Regarding the other parts of our framework, the optimizer and the predictive models are executed offline so their complexity does not affect the system. Hence, we can see that the complexity of GNCA depends on the number of videos and regions. In this context, increasing the number of cloud sites $R$ may impact the complexity of the algorithm. However, in real world scenarios, the current number of AWS cloud sites is 22, the number of Facebook data centers is 12, 20 for Google and 54 for Microsoft, which is considered as acceptable when being implemented in high computational servers. Furthermore, this heuristic is run, each time a small batch of videos $V$ is received. In this way, videos are allocated on-the-fly. To summarize, with the same complexity, our proposed heuristic GNCA outperforms the benchmarking baselines and the recent work GMC, owing to the high accuracy of the models that forecast the optimal number of required resources to rent with lower prices. Moreover, thanks to our strategy that chooses the nearest cloud sites in case resources are available and the cheapest in case resources need to be rented on-demand, GNCA achieved a good trade-off between cost and latency. 
\section{Conclusion}\label{section:conclusion}
In this paper, we proposed a novel geo-distributed computational resource allocation framework based on proactive reservation. First, our crowdsourcing videos allocation on geo-distributed platform is formulated as an optimization. This optimization is used offline on historical live videos to construct our timeseries datasets for different cloud regions. Second, the datasets are trained to forecast the number of cloud instances required for the upcoming time slot. Finally, a heuristic, namely GNCA, is designed to use the rented resources and allocate the live videos on-the-fly, while maximizing the QoS. For future work, We plan
to deal with the resource allocation problem using the Reinforcement Learning approach that is perfectly adequate for dynamic systems, and evaluate its performance against our proposed GNCA heuristic. 
\bibliographystyle{IEEEtran}
\bibliography{final_version.bib}
\begin{IEEEbiography}[{\includegraphics[width=1in,height=1.3in,clip]{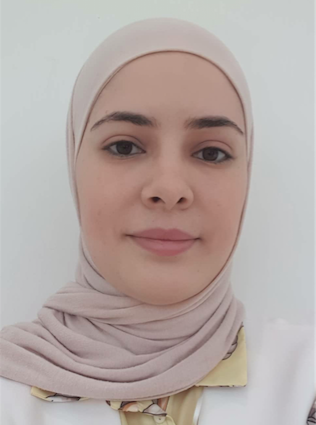}}]{Emna Baccour}
received the Ph.D. degree in computer Science from the University of Burgundy, France, in 2017. She was a postdoctoral fellow at Qatar University on a project covering the interconnection networks for massive data centers and then on a project covering video caching and processing in mobile edge computing networks. She currently holds a postdoctoral position at Hamad Ben Khalifa University. Her research interests include data center networks, cloud computing, green computing and software defined networks as well as distributed systems. She is also interested in edge networks and mobile edge caching and computing.
\end{IEEEbiography}
\vskip -1\baselineskip plus -1fil
\begin{IEEEbiography}[{\includegraphics[width=1in,height=1.3in,clip]{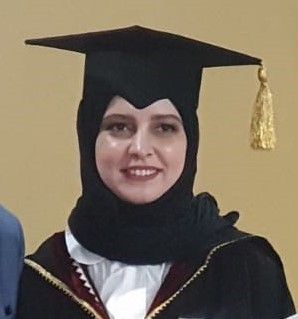}}]{Fatima Haouari}
received the B.Sc. and MSc. degree in computer science (with distinction) from Qatar University. She was a Research Assistant at Qatar university on a project focusing on Cloud-Enabled interactive multimedia applications for the crowd. She is currently a Ph.D. student at Qatar University working on early detection of fake news on social media. Her research interests include information retrieval and natural language processing. She is also interested in big data analytics and distributed systems.
\end{IEEEbiography}
\vskip -1\baselineskip plus -1fil
\begin{IEEEbiography}[{\includegraphics[width=1in,height=1.3in,clip]{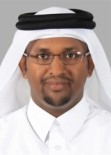}}]{Aiman Erbad}
is an Associate Professor at the College of Science and Engineering at Hamad Bin Khalifa University (HBKU). Dr. Erbad obtained a PhD in Computer Science from the University of British Columbia (Canada), and a Master of Computer Science in Embedded Systems and Robotics from the University of Essex (UK). Dr. Erbad received the Platinum award from H.H. The Emir Sheikh Tamim bin Hamad Al Thani at the Education Excellence Day 2013 (PhD category). Dr. Erbad received the 2020 best research paper award from the Computer Communications journal, IWCMC 2019 best paper award, and IEEE CCWC 2017 best paper award. Dr. Erbad is an editor in KSII Transactions on Internet and Information Systems and was a guest editor in IEEE Networks. Dr. Erbad research interests span cloud computing, edge computing, IoT, private and secure networks, and multimedia systems.
\end{IEEEbiography}
\vskip -1\baselineskip plus -1fil
\begin{IEEEbiography}[{\includegraphics[width=1.1in,height=1.3in,clip]{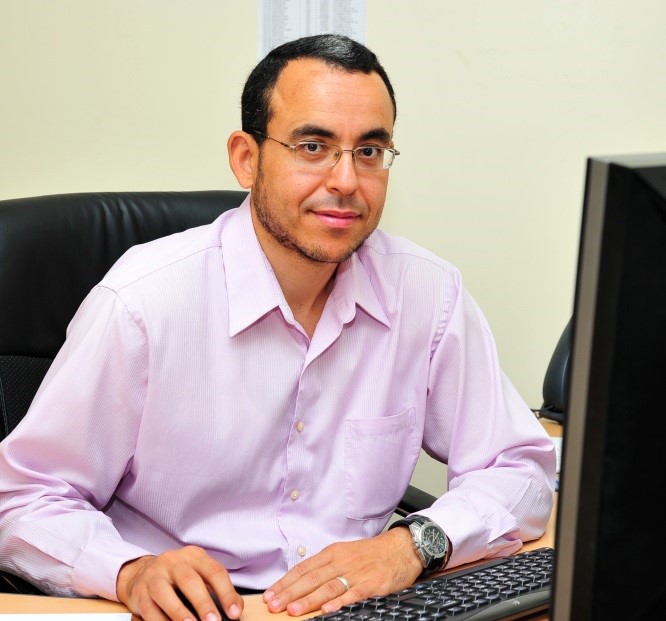}}]{Amr Mohammed}
(S’ 00, M’ 06, SM’ 14) received his M.S. and Ph.D. in electrical and computer engineering from the University of British Columbia, Vancouver, Canada, in 2001, and 2006 respectively. He has worked as an advisory IT specialist in IBM Innovation Centre in Vancouver from 1998 to 2007, taking a leadership role in systems development for vertical industries.  He is currently a professor in the college of engineering at Qatar University and the director of the Cisco Regional Academy. He has over 25 years of experience in wireless networking research and industrial systems development. He holds 3 awards from IBM Canada for his achievements and leadership, and 4 best paper awards from IEEE conferences. His research interests include wireless networking, and edge computing for IoT applications. Dr. Amr Mohamed has authored or co-authored over 160 refereed journal and conference papers, textbook, and book chapters in reputable international journals, and conferences. He is serving as a technical editor for the journal of internet technology and the international journal of sensor networks. He has served as a technical program committee (TPC) co-chair for workshops in IEEE WCNC’16. He has served as a co-chair for technical symposia of international conferences, including Globecom’16, Crowncom’15, AICCSA’14, IEEE WLN’11, and IEEE ICT’10. He has served on the organization committee of many other international conferences as a TPC member, including the IEEE ICC, GLOBECOM, WCNC, LCN and PIMRC, and a technical reviewer for many international IEEE, ACM, Elsevier, Springer, and Wiley journals.
\end{IEEEbiography}
\vskip -1\baselineskip plus -1fil
\begin{IEEEbiography}[{\includegraphics[width=1in,height=1.3in,clip]{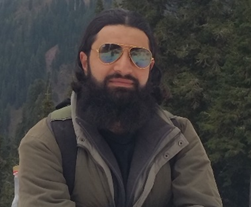}}]{Kashif Bilal}
received his PhD from North Dakota State University USA. He is an Assistant Professor at COMSATS University Islamabad, Abbottabad, Pakistan. His research interests include Edge computing, cloud, energy efficient high speed networks, and crowdsourced multimedia. Kashif is awarded North Dakota State University CoE Student Researcher of the year 2014 and COMSATS CS Researcher of the year in 2016 and 2020 based on his research contributions.
\end{IEEEbiography}
\vskip -1\baselineskip plus -1fil
\begin{IEEEbiography}[{\includegraphics[width=1in,height=1.3in,clip]{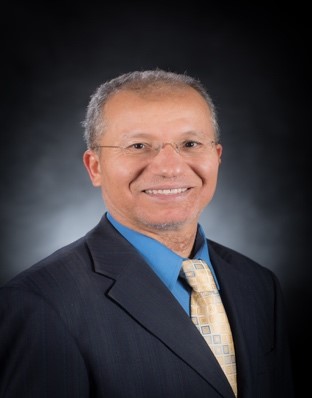}}]{Mohsen Guizani}
(S’85–M’89–SM’99–F’09) received the B.S. (with distinction) and M.S. degrees in electrical engineering, the M.S. and Ph.D. degrees in computer engineering from Syracuse University, Syracuse, NY, USA, in 1984, 1986, 1987, and 1990, respectively. He is currently a Professor at the Computer Science and Engineering Department in Qatar University, Qatar. Previously, he served in different academic and administrative positions at the University of Idaho, Western Michigan University, University of West Florida, University of Missouri-Kansas City, University of Colorado-Boulder, and Syracuse University. His research interests include wireless communications and mobile computing, computer networks, mobile cloud computing, security, and smart grid. He is currently the Editor-in-Chief of the IEEE Network Magazine, serves on the editorial boards of several international technical journals and the Founder and Editor-in-Chief of Wireless Communications and Mobile Computing journal (Wiley). He is the author of nine books and more than 500 publications in refereed journals and conferences. He guest edited a number of special issues in IEEE journals and magazines. He also served as a member, Chair, and General Chair of a number of international conferences. Throughout his career, he received three teaching awards and four research awards. He also received the 2017 IEEE Communications Society WTC Recognition Award as well as the 2018 AdHoc Technical Committee Recognition Award for his contribution to outstanding research in wireless communications and Ad-Hoc Sensor networks. He was the Chair of the IEEE Communications Society Wireless Technical Committee and the Chair of the TAOS Technical Committee. He served as the IEEE Computer Society Distinguished Speaker and is currently the IEEE ComSoc Distinguished Lecturer. He is a Fellow of IEEE and a Senior Member of ACM.
\end{IEEEbiography}
\vskip -1\baselineskip plus -1fil
\begin{IEEEbiography}[{\includegraphics[width=1in,height=1.3in,clip]{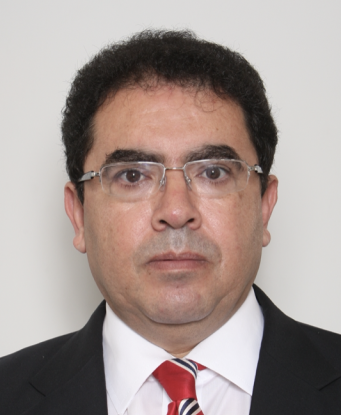}}]{Mounir Hamdi}
received the B.S. degree (Hons.) in electrical engineering (computer engineering) from the University of Louisiana, in 1985, and the M.S. and Ph.D. degrees in electrical engineering from the University of Pittsburgh, in 1987 and 1991, respectively. He was a Chair Professor and a Founding Member of The Hong Kong University of Science and Technology (HKUST), where he was the Head of the Department of Computer Science and Engineering. From 1999 to 2000, he held visiting professor positions at Stanford University and the Swiss Federal Institute of Technology. He is currently the Founding Dean of the College of Science and Engineering, Hamad Bin Khalifa University (HBKU). His area of research is in high-speed wired/wireless networking, in which he has published more than 360 publications, graduated more 50 M.S./Ph.D. students, and awarded numerous research grants. In addition, he has frequently consulted for companies and governmental organizations in the USA, Europe, and Asia. He is a Fellow of the IEEE for his contributions to design and analysis of high-speed packet switching, which is the highest research distinction bestowed by IEEE. He is also a frequent keynote speaker in international conferences and forums. He is/was on the editorial board of more than ten prestigious journals and magazines. He has chaired more than 20 international conferences and workshops. In addition to his commitment to research and academic/professional service, he is also a dedicated teacher and a quality assurance educator. He received the Best 10 Lecturer Award and the Distinguished Engineering Teaching Appreciation Award from HKUST. He is frequently involved in higher education quality assurance activities as well as engineering programs accreditation all over the world.
\end{IEEEbiography}
\end{document}